\documentclass[11pt]{article}
\usepackage{amsmath, amssymb, graphicx, cite, hyperref}
\usepackage{geometry}
\usepackage{tikz}
\usepackage{authblk}
\usepackage{enumitem}
\usepackage{array}
\usepackage{booktabs}
\usepackage{float}           % For the [H] placement option
\usepackage{subcaption}      % For subfigure environments
\usepackage{graphicx}        % For including graphics

\usetikzlibrary{shapes.geometric, arrows}

\tikzstyle{startstop} = [rectangle, rounded corners, minimum width=3cm, minimum height=1cm, text centered, draw=black, fill=red!30]
\tikzstyle{process} = [rectangle, minimum width=3cm, minimum height=1cm, text centered, draw=black, fill=blue!20]

\title{Physics-Informed Neural Networks in Clean Combustion: A Pathway to Sustainable Aerospace Propulsion}
\author[1]{Caleb Caldwell}
\author[1]{Jacob Baltes}
\author[2,*]{Mahmood Mousavi}
\author[1]{Muteb Aljasem}
\author[3]{Bok Jik Lee}

\affil[1]{Department of Robotics, Electronics, and Computer Engineering, Bowling Green State University, Bowling Green, OH, USA}
\affil[2]{School of Engineering, Embry-Riddle Aeronautical University - Worldwide, Daytona Beach, FL 32114, USA}
\affil[3]{Department of Aerospace Engineering, Seoul National University, Seoul, 08826, Republic of Korea}
\date{} % Removes the date

\begin{document}

\maketitle

\begin{abstract}

Achieving clean combustion systems is crucial in terms of solving environmental impacts, decarbonization needs and sustainability matters. Traditional combustion modeling techniques via computational fluid dynamics with accurate chemical kinetics face obstacles in computational cost and accurate representation of turbulence-chemistry interactions. Physically Informed Neural Networks (PINNs) as a new framework, merges physical laws with data-driven learning and shows great potential as an alternative methodology. By directly integrating conservation equations into their training process, PINNs achieve accurate mesh-free modeling of complex combustion phenomena despite having limited data sets. This review examines how this approach applies to clean combustion systems while focusing on their impact in aerospace applications including flame dynamics, turbulent combustion, emission prediction, and instability management in propulsion systems. Next-generation aerospace engines rely on PINNs to reduce computational costs while increasing predictive performance and enabling real-time control methods. This analysis concludes by exploring current barriers and future paths, while demonstrating how PINNs can revolutionize sustainable and efficient combustion technologies in aerospace propulsion systems.

\end{abstract}

\noindent\textbf{Keywords:} Physics-Informed Neural Networks; Clean Combustion;  Data-Driven Modeling

\section{Introduction}

The increasing demand for sustainable energy solutions and the urgent need to mitigate climate change have intensified the focus on clean combustion technologies, which aim to reduce emissions, improve fuel efficiency, and support the transition to renewable energy sources \cite{turns1996introduction, glassman2014combustion}. While computational fluid dynamics (CFD) and chemical kinetics solvers have provided valuable insights, they face challenges such as high computational costs, turbulence-chemistry interaction complexities, and the need for extensive empirical models \cite{poinsot2005theoretical, warnatz2006combustion}.

Despite advancements, traditional models struggle to capture turbulence-chemistry interactions and require significant computational resources. Additionally, limited high-fidelity experimental data hinders model validation and refinement, leaving gaps in our understanding of clean combustion.

Clean combustion encompasses technologies such as lean-burn and oxy-fuel combustion, as well as alternative fuels like hydrogen and ammonia. However, the interplay of fluid dynamics, chemical kinetics, and heat transfer presents significant challenges for conventional modeling techniques. Addressing these challenges requires accurate, efficient, and robust predictive models.

The rise of artificial intelligence (AI) and machine learning (ML) has introduced new possibilities for combustion research. Physics-informed neural networks (PINNs) integrate physical laws, formulated as partial differential equations (PDEs), into their learning process, ensuring adherence to fundamental physics while reducing reliance on large datasets \cite{raissi2019physics, karniadakis2021physics}. Unlike traditional neural networks, PINNs embed governing equations into the loss function, improving predictive accuracy and generalizability.

PINNs offer a flexible framework for solving forward and inverse problems in combustion, providing accurate simulations with lower computational costs. Their ability to capture nonlinear dynamics makes them particularly effective in applications such as flame dynamics, turbulent combustion modeling, emissions prediction, and chemical kinetics \cite{cuomo2022scientific}. 

PINNs have demonstrated promising results in modeling laminar and turbulent flames, capturing key features such as flame speed, stretch effects, and extinction limits with high accuracy \cite{zhang2022physics}. Additionally, they have been applied to emissions prediction, optimizing combustion systems to meet stringent environmental regulations \cite{sha2025physics, zhang2024model}. Their capability for real-time predictions enhances operational efficiency and reduces emissions \cite{zhang2024crk, luo2025exploring}.

This review provides a comprehensive overview of the application of PINNs in clean combustion, highlighting how they address limitations in traditional models, particularly in handling turbulence-chemistry interactions, reducing computational costs, and improving predictive capabilities. We also discuss the current challenges and future perspectives of PINNs in advancing clean combustion technologies. By bridging the gap between conventional models and modern AI techniques, this review underscores the transformative potential of PINNs in promoting sustainable energy solutions.

\section{Fundamentals of Physics-Informed Neural Networks}

PINNs represent a transformative approach in the realm of computational modeling, blending the data-driven prowess of machine learning with the rigorous foundations of physical laws. The concept of PINNs revolves around embedding PDEs, which govern a wide range of physical phenomena, directly into the structure of neural networks. This integration allows for the learning of solutions to complex problems without the exhaustive need for traditional numerical discretization techniques.
\begin{figure}[h]
    \centering
    \includegraphics[width=0.9\textwidth]{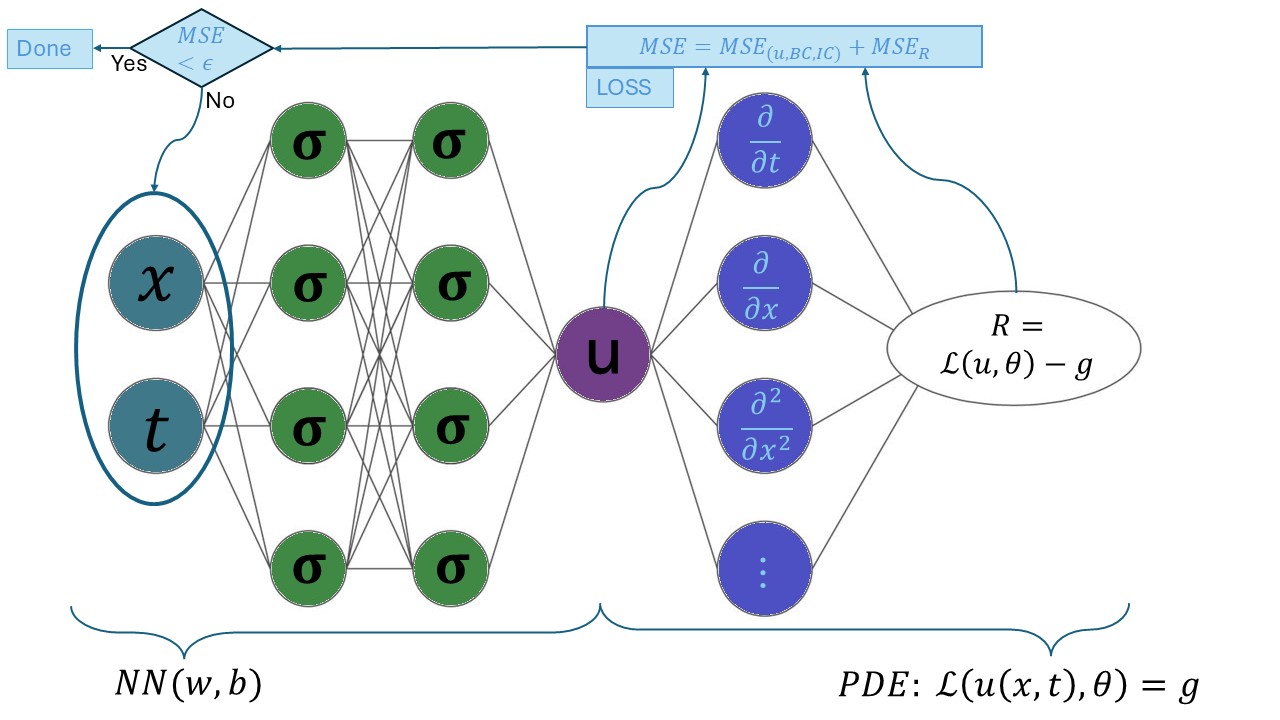}
    \caption{Schematic diagram of a PINNs. The neural network takes spatial and temporal variables as inputs and is trained to satisfy both the data and the governing physical laws, represented by the PDE residuals. \cite{meng2020ppinn}}
    \label{fig:PINN_Struc}
\end{figure}

As illustrated in Fig.~\ref{fig:PINN_Struc}, a PINNs is designed to accept inputs such as spatial and temporal variables. The network is trained not only to fit observed data but also to minimize the residuals of the governing PDEs, ensuring that the learned solutions adhere to the underlying physical laws.

\subsection{Architecture and Mathematical Formulation}

The architecture of PINNs is typically composed of deep feedforward neural networks designed to approximate the solution to the governing equations of a given physical system. The input layer accepts spatial and temporal coordinates \( \mathbf{x} \) and \( t \), while the output layer provides the predicted physical quantities, such as temperature, velocity, or species concentration. Hidden layers with nonlinear activation functions facilitate the network's ability to capture complex patterns and relationships inherent in the data.

A distinctive feature of PINNs is their loss function, which is formulated to include both data-driven components and physics-based constraints. This composite loss function comprises terms that penalize deviations from known data points and terms that enforce the satisfaction of the governing PDEs, boundary conditions, and initial conditions. The general form of the governing PDE can be written as:

\begin{equation}
\mathcal{N}[u(\mathbf{x},t)] = 0, \quad \mathbf{x} \in \Omega, \; t \in [0, T],
\end{equation}

where \( \mathcal{N} \) represents a nonlinear differential operator, \( u(\mathbf{x},t) \) is the solution function, and \( \Omega \) denotes the spatial domain. The boundary and initial conditions are defined as:

\begin{equation}
\mathcal{B}[u(\mathbf{x},t)] = g(\mathbf{x},t), \quad \mathbf{x} \in \partial\Omega, \; t \in [0, T],
\end{equation}

\begin{equation}
 u(\mathbf{x},0) = u_0(\mathbf{x}), \quad \mathbf{x} \in \Omega.
\end{equation}

The loss function for PINNs integrates these conditions as follows:

\begin{equation}
\mathcal{L}_{\text{total}} = \mathcal{L}_{\text{data}} + \lambda_{\text{PDE}}\mathcal{L}_{\text{PDE}} + \lambda_{\text{BC}}\mathcal{L}_{\text{BC}} + \lambda_{\text{IC}}\mathcal{L}_{\text{IC}},
\end{equation}

where \( \lambda \) are weighting factors for each constraint, ensuring balanced learning across data fidelity and physical accuracy.

\subsection{PINNs Algorithm}

The algorithmic framework of PINNs is illustrated in the flowchart shown in Fig.~\ref{fig:pinn_flowchart}. It outlines the sequential steps typically followed in the development and deployment of a PINNs model:

\tikzstyle{arrow} = [thick,->,>=stealth]
\begin{figure}[ht]
\centering
\begin{center}
\begin{tikzpicture}[node distance=1.5cm]
    % Define nodes with simpler text
    \node (start) [startstop] {\textbf{Start: Define Problem and}\par\textbf{Governing PDEs}};
    \node (init) [process, below of=start] {\textbf{Step 1: Initialize Neural}\par\textbf{Network Architecture}};
    \node (form) [process, below of=init] {\textbf{Step 2: Formulate}\par\textbf{Composite Loss Function}};
    \node (train) [process, below of=form] {\textbf{Step 3: Train Network}\par\textbf{Using Optimization}};
    \node (validate) [process, below of=train] {\textbf{Step 4: Validate Model}\par\textbf{with Data}};
    \node (deploy) [process, below of=validate] {\textbf{Step 5: Deploy Model}\par\textbf{for Predictions}};
    \node (end) [startstop, below of=deploy] {\textbf{End: Application in}\par\textbf{Real-World Scenarios}};

    % Draw arrows
    \draw [arrow] (start) -- (init);
    \draw [arrow] (init) -- (form);
    \draw [arrow] (form) -- (train);
    \draw [arrow] (train) -- (validate);
    \draw [arrow] (validate) -- (deploy);
    \draw [arrow] (deploy) -- (end);
\end{tikzpicture}
\end{center}
\caption{Algorithmic framework of PINNs. Adapted from Liu et al. \cite{liu2024adaptive}.}
\label{fig:pinn_flowchart}
\end{figure}
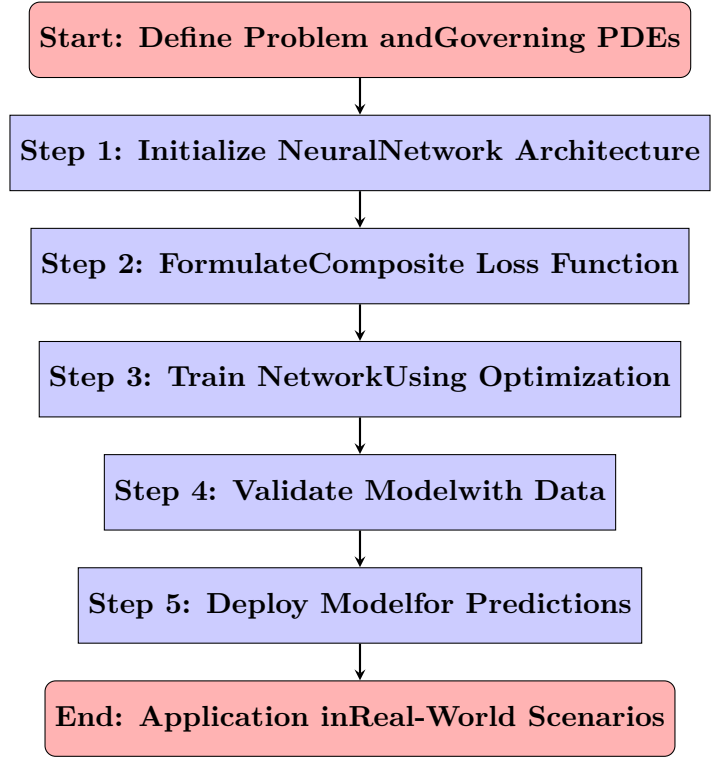

\subsection{Advantages of PINNs in Combustion Modeling}
PINNs present several advantages over traditional computational approaches for combustion modeling. Conventional methods, including finite element and finite volume techniques, typically require extensive mesh generation. This process is time-consuming and computationally expensive, particularly for complex geometries found in internal combustion engines or for simulating transient, multidimensional flame propagation. In contrast, PINNs offer a mesh-free modeling framework, significantly reducing computational complexity and enabling faster model development and analysis \cite{taassob2023physics,cuomo2022scientific,singh2024physics,chen2024physics,farea2024understanding}.

For example, Jeon et al.\cite{jeon2022finite} developed a hybrid approach combining finite volume methods (FVM) with a PINNs. Their work demonstrated computational speeds nearly ten times faster than traditional CFD solvers in reacting flow datasets, emphasizing the efficiency and scalability of PINNs.

The dual capability of PINNs to solve both forward and inverse problems is another key advantage. In forward modeling, PINNs predict system behaviors based on known parameters, while in inverse problems, they infer unknown parameters such as reaction rates and transport properties from observed data \cite{raissi2019physics}. This is exemplified by the FlamePINN-1D framework, which effectively addresses both forward and inverse problems for one-dimensional laminar flames \cite{wu2025flamepinn}.

Moreover, PINNs generalize well from limited datasets by leveraging embedded physical laws. This feature is crucial in combustion applications where acquiring experimental data is often difficult due to the extreme operating conditions. Studies have shown that PINNs can maintain high predictive accuracy even in data-scarce environments \cite{ganga2024exploring, bonfanti2302generalization, deresse2024exploring, schafer2022generalization, mishra2023estimates}.

By integrating data-driven modeling with fundamental physical principles, PINNs improve the reliability, efficiency, and interpretability of combustion simulations. These advantages make them a promising tool for advancing combustion research and optimizing clean energy systems \cite{jagtap2022physics}.

\subsection{Challenges and Limitations}

While PINNs offer substantial benefits in combustion modeling, they also present notable challenges when applied to complex physical systems. One major difficulty lies in optimizing the composite loss function, particularly for stiff PDEs and highly nonlinear dynamics common in combustion. Competing objectives within the loss terms can lead to gradient imbalance, where certain constraints dominate the optimization and hinder convergence \cite{bonfanti2024challenges, basir2022investigating, anantharaman2023approximation}.

Training PINNs for high-dimensional or multi-scale problems can be computationally demanding, often requiring considerable time and resources. Additionally, their performance is highly sensitive to neural network architecture and hyperparameter choices \cite{hu2024tackling}. Unlike traditional CFD methods with standardized practices, PINNs frequently rely on empirical design and manual tuning, which can be time-consuming and inconsistent \cite{el2024importance}.

Another challenge is their limited ability to capture sharp gradients, discontinuities, and localized phenomena—characteristic features in combustion systems like shock waves and flame fronts. These limitations can compromise solution accuracy in regions with strong nonlinearities or abrupt transitions.

However, adaptive learning rates \cite{jagtap2020locally}, Neural Architecture Search (NAS) \cite{wang2022auto}, and domain decomposition techniques such as Separable PINNs (SPINN) \cite{cho2024separable} have shown promise in improving convergence, reducing computational overhead, and enhancing accuracy.

Furthermore, integrating noisy experimental data introduces additional complexities. Data imperfections can degrade model performance, necessitating advanced regularization methods and robust preprocessing pipelines \cite{peng2022robust, pilar2024physics, xu2024preprocessing}.

Despite these challenges, PINNs remain a promising tool for advancing combustion modeling. Ongoing research efforts aim to overcome current limitations, improving their reliability and expanding their applicability to complex reactive flows.

\section{Clean Combustion: An Overview}

Clean combustion is defined as the process of burning fuels in a way that minimizes pollutant emissions, increases fuel efficiency, and contributes to environmental protection \cite{battin2013cleaner}. The main focus is on ensuring complete and efficient combustion of the fuel-air mixture, reducing the generation of harmful byproducts such as nitrogen oxides (NO\textsubscript{x}), carbon monoxide (CO), and particulate matter. At the same time, maximizing energy conversion efficiency remains a primary goal \cite{glassman2014combustion, warnatz2006combustion}.

Achieving high thermal efficiency is a fundamental objective of clean combustion. This is accomplished by optimizing critical combustion parameters such as temperature, pressure, and fuel-air equivalence ratio \cite{szybist2021fuel}. Effective control of these parameters promotes complete oxidation of the fuel, reduces unburned hydrocarbons, and limits energy losses during the combustion process \cite{mousavi2024effects}.

Another major aspect of clean combustion is emissions control. Reducing pollutants is crucial for meeting environmental regulations and minimizing health impacts \cite{mousavi2019comprehensive, mousavi2022effects}. Technologies such as advanced burner designs, exhaust gas recirculation (EGR), and catalytic converters are commonly used to control and reduce emissions \cite{turns1996introduction}.

Sustainable energy practices are integral to clean combustion initiatives. The use of alternative fuels—including hydrogen, ammonia, and biofuels—not only reduces dependency on fossil fuels but also helps lower greenhouse gas emissions \cite{kabeyi2022sustainable, mansoori2021fuels}. Additionally, advanced combustion control systems with real-time monitoring and feedback capabilities enable dynamic adjustment of operating conditions, ensuring consistent performance and minimizing emissions under varying loads and fuel compositions \cite{stanvcin2020review}.

\subsection{Hydrogen Combustion}

Hydrogen combustion is considered a clean fuel option due to its high energy content and zero carbon emissions when combusted. Hydrogen combustion produces water vapor as the primary byproduct, making it an attractive option for sustainable energy systems \cite{nowotny2011impact, habib2024hydrogen}. However, challenges such as high flame speeds, the risk of pre-ignition, and NO\textsubscript{x} formation at high temperatures require careful management. Innovative techniques, such as staged combustion and the use of diluents, are employed to mitigate these issues and enhance combustion stability \cite{elbaz2019low, chun2023effects, zhang2015moderate, mousavi2019comprehensive}.

\subsection{Ammonia Combustion}

Ammonia (NH\textsubscript{3}) is gaining attention as a carbon-free fuel with potential applications in power generation and transportation. Ammonia combustion can produce nitrogen and water, but it also poses challenges related to low reactivity, incomplete combustion, and NO\textsubscript{x} emissions \cite{mousavi2024effects, alnajideen2024ammonia, valera2024ammonia, kobayashi2019science}. Advanced combustion strategies, including the use of catalysts and co-firing with hydrogen, are being explored to improve ammonia combustion characteristics and reduce its environmental impact \cite{kang2023review, mousavi2022effects}.

\subsection{Syngas and Biofuels}

Synthetic gas (syngas) and biofuels offer renewable alternatives to conventional fossil fuels \cite{rozzi2020green, cherwoo2023biofuels}. Syngas, a mixture of hydrogen, carbon monoxide, and other gases, can be produced from biomass, coal, or natural gas. Biofuels, derived from biological sources, provide a sustainable energy solution with reduced greenhouse gas emissions \cite{deora2022biofuels, ridjan2013feasibility}. The combustion of these fuels involves complex chemical kinetics and requires optimized combustion systems to ensure efficiency and low emissions. Innovations in burner design and fuel injection techniques are critical for enhancing the performance of syngas and biofuels in various combustion applications \cite{dinccer2012fossil}.

\subsection{Oxy-Fuel and Lean-Burn Combustion}

Oxy-fuel combustion involves burning fuel in pure oxygen instead of air, resulting in higher flame temperatures and the production of a concentrated CO\textsubscript{2} stream suitable for carbon capture and storage \cite{nemitallah2017oxy, toftegaard2010oxy, scheffknecht2011oxy, habib2019oxyfuel}. This technology offers significant advantages for reducing greenhouse gas emissions but also presents challenges related to heat management and material durability. Lean-burn combustion operates with excess air, reducing peak temperatures and NO\textsubscript{x} emissions \cite{raho2025technological}. Both technologies contribute to cleaner combustion processes but require advanced control mechanisms to maintain stability and efficiency.

\subsection{Challenges in Modeling Clean Combustion Processes}

Modeling clean combustion processes is inherently complex due to the intricate coupling between fluid dynamics, chemical kinetics, and heat transfer. Clean combustion typically involves multi-step reaction mechanisms with numerous intermediate species and reaction pathways, making accurate simulation highly challenging \cite{warnatz2006combustion}. Capturing the interaction between turbulent flow fields and chemical reactions is particularly difficult, yet it is crucial for predicting flame structure, propagation speed, and pollutant formation \cite{poinsot2005theoretical}.

The accurate prediction of NO\textsubscript{x} emissions and particulate matter formation hinges on high-fidelity models that can resolve intricate turbulence-chemistry interactions. Traditional CFD methods often struggle to capture these complexities, primarily due to their high computational demands and the need to address phenomena occurring across multiple spatial and temporal scales \cite{turns1996introduction}. PINNs provide a scalable and computationally efficient framework for modeling complex combustion systems with enhanced accuracy \cite{raissi2019physics, karniadakis2021physics}.

\section{Applications of PINNs in Clean Combustion}
\subsection{Flame Dynamics and Propagation}
Understanding flame dynamics and propagation is vital in combustion science, directly influencing efficiency, stability, and emission control. Traditional models predict flame behavior by solving complex PDEs that govern mass, momentum, and energy conservation, often coupled with detailed chemical kinetics. However, these methods struggle to capture nonlinear turbulence-chemistry interactions, particularly under the extreme conditions of clean combustion systems \cite{poinsot2005theoretical, warnatz2006combustion}.
PINNs address these challenges by accurately modeling flame dynamics, even when data is limited or incomplete. They approximate PDE solutions while ensuring consistency with governing physical laws \cite{raissi2019physics, karniadakis2021physics}.

A key parameter in flame dynamics is flame speed, critical for optimizing engine performance and reducing emissions. PINNs have successfully modeled both laminar and turbulent flame speeds, capturing dependencies on pressure, temperature, and equivalence ratio. Zhang et al. \cite{zhang2024crk} demonstrated that CRK-PINNs achieved 6.0-14.6 times faster chemical source term computations and 2.3-4.9 times overall simulation speedups compared to direct integration methods. These improvements were validated across cases including 0-D autoignition, 2-D Bunsen flames, and 3-D turbulent jet flames, with data generation and training representing just 2.68\% of total simulation time.

Flame stretch, arising from strain and curvature, significantly impacts flame stability and extinction \cite{choi2000flame}. Traditional approaches rely on detailed flow field data and complex simulations, while PINNs offer more efficient solutions by its approach. This is valuable in lean and diluted clean combustion applications where the stability of the flame is more difficult to maintain \cite{taassob2023physics, wu2025flamepinn}.

Extinction limits define the conditions under which a flame can no longer sustain combustion \cite{MATALON200957}. These limits are governed by factors such as heat loss, strain, and reaction kinetics \cite{lee2024large}. PINNs enable accurate prediction of extinction limits, facilitating the identification of stability thresholds essential for optimized combustion system design \cite{Song11}.

PINNs have been applied in both premixed and non-premixed flame studies. In premixed flames, they model flame front propagation by balancing chemical reactions and fluid dynamics. In non-premixed cases, they offer insights into flame structure, ignition dynamics, and pollutant formation. Across these applications, PINNs demonstrate accuracy, efficiency, and flexibility that often surpass conventional techniques \cite{cuomo2022scientific}.

Table~\ref{tab:comparison} compares PINNs and traditional CFD models in flame dynamics. While CFD offers slightly higher accuracy in some scenarios, PINNs excel in computational efficiency and scalability, making them well-suited for real-time applications.

\begin{table}[h!]
\centering
\caption{Comparative Performance of Traditional CFD Models and PINNs in Flame Dynamics}
\begin{tabular}{p{3cm} p{5cm} p{5cm}}
\hline
\textbf{Aspect} & \textbf{Traditional CFD} & \textbf{PINNs} \\
\hline
\textbf{Accuracy} & High accuracy in simulating complex fluid dynamics; may require fine meshes and small time steps, leading to increased computational costs. & Capable of achieving comparable accuracy by incorporating physical laws into the neural network training process. \\

\textbf{Computational Cost} & Computationally intensive, especially for high-fidelity simulations involving detailed chemistry and turbulence. & Potentially more efficient by leveraging neural networks, but training can be time-consuming and may require substantial data. \\

\textbf{Scalability} & Scalability can be challenging due to the need for mesh generation and solver adjustments for different geometries and conditions. & Offers high scalability, particularly in parametric studies, as the trained model can generalize to different scenarios without retraining. \\

\textbf{Data Dependency} & Relies on numerical discretization and does not require experimental data for simulations. & Requires data for training; the quality and quantity of data can significantly impact performance. \\

\textbf{Implementation} & Established methodologies with numerous validated models and solvers; however, setup can be complex and time-consuming. & Emerging approach with increasing research interest; implementation may require expertise in both physics and machine learning. \\
\hline
\end{tabular}
\label{tab:comparison}
\end{table}

\subsection{Turbulent Combustion Modeling}
Turbulent combustion involves complex interactions between turbulent flow and chemical reactions, critically impacting flame stability, heat release, and pollutant formation. Accurately modeling these turbulence-chemistry interactions remains challenging due to their nonlinear, multi-scale nature \cite{pope2001turbulent}.

PINNs, with their inherent nonlinear architecture and ability to integrate physical laws within neural network training, have emerged as a powerful approach for modeling turbulent combustion \cite{taassob2024pinn}. 

Recent studies demonstrate the effectiveness of PINNs in capturing the intricate dynamics of turbulent combustion. For example, a PINNs framework was developed to evaluate closure terms for turbulence and chemical source terms in turbulent non-premixed flames. By leveraging temperature, species concentration, and velocity measurements, the model accurately reconstructed principal components and velocity fields, showing strong agreement with experimental data \cite{taassob2023physics}.

PINNs have also been utilized to investigate thermoacoustic interactions in combustors, which play a crucial role in combustion instability analysis. By integrating acoustic pressure measurements and heat release rate data, PINN-based models successfully captured the coupled dynamics of vortex shedding and acoustic oscillations \cite{mariappan2024learning, xie2024predicting}.

A notable advancement in this field is the development of a Dual-Path neural network designed to model the nonlinear thermoacoustic response of flames in the time domain \cite{wu2024dual}. This architecture introduces two distinct learning paths: the Chronological Feature Path (CFP) for capturing time-sequenced features and the Temporal Detail Feature Path (TDFP) for learning fine temporal details. Unlike conventional high-fidelity numerical simulations, which are computationally demanding, this model leverages frequency-sweeping datasets to accurately reconstruct flame responses using limited input data. The Dual-Path approach demonstrates superior performance in predicting nonlinear flame responses, particularly in strongly nonlinear regimes.

\subsection{Emissions Prediction and Control}
Accurate prediction and control of pollutant emissions—such as NO\textsubscript{x}, soot, and CO—are critical for developing clean combustion systems that meet stringent environmental regulations. Traditional CFD approaches for modeling pollutant formation rely on detailed chemical kinetics and high-fidelity simulations. While these methods provide valuable insights, they are computationally intensive and often depend on large datasets.

To overcome these challenges, PINNs have emerged as efficient alternatives, offering high prediction accuracy while significantly reducing computational costs. PINNs integrate physical constraints with data-driven learning, minimizing the need for extensive training datasets and improving generalization.

For example, a physics-informed deep learning framework was developed to predict NO\textsubscript{x} concentrations by encoding advection-diffusion mechanisms and fluid dynamic constraints directly into the network. This approach effectively reduced model bias and enhanced predictive performance \cite{li2023physics}. Similarly, Cao et al. \cite{cao2024physics} demonstrated that incorporating domain knowledge through PDEs enables PINNs to maintain high fidelity in NO\textsubscript{x} emission predictions, even with limited data availability.

In industrial-scale applications, such as coal-fired power plants, PINNs have shown considerable promise. Zhu et al. \cite{zhu2024physics} introduced a PINNs model with monotonicity constraints to predict NO\textsubscript{x} emissions at the boiler outlet. Their results indicated superior accuracy and generalization compared to conventional machine learning models, particularly under varying operational conditions.

Soot formation prediction has also benefited from PINN-based methods. Traditional approaches require solving complex chemical kinetics and transport equations, making them computationally demanding. PINNs offer a reduced-order modeling framework that efficiently captures the essential physics and chemistry governing soot formation \cite{liu2024reconstructing, wang2024soot}. In soot-laden flames, where laser diagnostics are limited, PINNs have been successfully employed to reconstruct velocity and temperature fields, ensuring physically consistent results. Moreover, Liu et al. \cite{liu2024reconstructing} proposed an Autoencoder-based Deep Operator Network (AE-DeepONet) for soot volume fraction prediction, achieving an 80\% reduction in mean squared error (MSE) compared to conventional machine learning techniques.

Additionally, PINNs effectively model NO\textsubscript{x} formation by accurately capturing the strong temperature dependence characteristic of the extended Zeldovich mechanism, which dominates thermal NO\textsubscript{x} production \cite{cao2024physics}:
\begin{equation}
O + N_2 \rightleftharpoons NO + N
\end{equation}
\begin{equation}
N + O_2 \rightleftharpoons NO + O
\end{equation}
\begin{equation}
N + OH \rightleftharpoons NO + H
\end{equation}

Furthermore, Table~\ref{tab:nox_models} compares various machine learning models applied to NO\textsubscript{x} emission prediction in coal-fired boilers. PINNs consistently deliver superior accuracy and generalization, achieving test set $R^2$ values up to 0.96, outperforming traditional Deep Neural Networks (DNN), Random Forest, and Least-Squares Support Vector Regression (LSSVR) models \cite{zhu2024physics, li2023physics}.

While conventional models such as Random Forest and LSSVR perform reasonably well on training datasets, they often exhibit poor generalization under different operating conditions. The inclusion of physical constraints in PINN-based models not only improves prediction accuracy but also ensures physical consistency and interpretability. However, this comes at the expense of increased computational effort during training \cite{zhu2024physics, li2023physics}.

\begin{table}[htbp]
    \centering
    \caption{Comparison of Machine Learning Models}
    \small  % Reduce font size
    \begin{tabular}{p{2cm}p{2cm}p{2cm}p{2cm}p{3cm}p{3cm}}
        \toprule
        \textbf{Model} & \textbf{Type} & \textbf{Training} & \textbf{Testing} & \textbf{Advantages} & \textbf{Limitations} \\
        \midrule
        Physics-Guided NN & Neural network with physical constraints & 
        $R^2 \approx 0.98$ \newline RMSE: 4.0-4.2 & 
        $R^2 \approx 0.96-0.97$ \newline RMSE: 5.2-5.5 & 
        \begin{itemize}[leftmargin=*,nosep]
            \item Preserves physical relationships
            \item High interpretability
            \item Most accurate predictions
        \end{itemize} & 
        \begin{itemize}[leftmargin=*,nosep]
            \item Longer training time
            \item More complex implementation
            \item Requires domain knowledge
        \end{itemize} \\
        \midrule
        DNN & Pure deep neural network & 
        $R^2 \approx 0.96$ \newline RMSE: 5.0–5.5 & 
        $R^2 \approx 0.93-0.94$ \newline RMSE: 6.8–7.0 & 
        \begin{itemize}[leftmargin=*,nosep]
            \item Good accuracy on training data
            \item Fast training
            \item Simple implementation
        \end{itemize} & 
        \begin{itemize}[leftmargin=*,nosep]
            \item Poor physical interpretability
            \item Limited generalization
        \end{itemize} \\
        \midrule
        LSSVR & Support vector regression & 
        $R^2 \approx 0.87-0.89$ \newline RMSE: 9.5–9.7 & 
        $R^2 \approx 0.89$ \newline RMSE: 9.7–10.0 & 
        \begin{itemize}[leftmargin=*,nosep]
            \item Handles nonlinear relationships
            \item Works well with limited data
        \end{itemize} & 
        \begin{itemize}[leftmargin=*,nosep]
            \item Poor generalization
            \item High computational cost
            \item Limited interpretability
        \end{itemize} \\
        \midrule
        Random Forest & Ensemble learning & 
        $R^2 \approx 0.95$  \newline RMSE: 5.8–6.0 & 
        $R^2 \approx 0.93$  \newline RMSE: 6.9–7.0 & 
        \begin{itemize}[leftmargin=*,nosep]
            \item Robust to outliers
            \item Handles mixed data types
            \item Fast prediction
        \end{itemize} & 
        \begin{itemize}[leftmargin=*,nosep]
            \item Limited physical interpretability
            \item Can overfit
            \item Poor extrapolation
        \end{itemize} \\
        \bottomrule
    \end{tabular}
    \label{tab:nox_models}
\end{table}

CO emissions, primarily a result of incomplete combustion, also pose challenges to accurate prediction. PINNs have shown promise in modeling CO formation and oxidation by capturing the coupling between temperature fields, mixing processes, and chemical kinetics \cite{almeldein2024accelerating}. A key reaction governing CO oxidation is as follows.

\begin{equation}
CO + OH \rightleftharpoons CO_2 + H
\end{equation}

Although this reaction appears simple, it occurs within a complex network involving radical species and intermediate reactions \cite{feilberg2002co+}. PINNs effectively model these interactions by considering the relevant chemical kinetics and fluid dynamics, improving prediction accuracy for CO emissions in varying combustion conditions.

\subsection{Combustion Instabilities and Dynamics}

Combustion instabilities represent a critical challenge in the design and operation of modern propulsion and power generation systems. These instabilities, particularly thermoacoustic oscillations, can significantly impact system performance, component life, and operational reliability \cite{liu2024instability}. Recent advances in PINNs have emerged as powerful tools for analyzing and predicting these complex phenomena, offering new perspectives on stability analysis and control strategies \cite{mariappan2024learning}.

\subsubsection{Thermoacoustic Instabilities and Physical Mechanisms}

Thermoacoustic instabilities arise from the intricate coupling between unsteady heat release and acoustic pressure oscillations within combustion chambers \cite{zhang2025heat}. The governing equation for acoustic pressure fluctuations in a combustion chamber captures this coupling through a wave equation with a source term, as shown in Eq.~\eqref{eq:acoustic_wave}:

\begin{equation}
\frac{\partial^2 p'}{\partial t^2} - c^2\nabla^2p' = (\gamma-1)\frac{\partial q'}{\partial t}
\label{eq:acoustic_wave}
\end{equation}

where \( p' \) represents pressure fluctuations, \( c \) is the speed of sound, \( \gamma \) is the specific heat ratio, and \( q' \) denotes heat release fluctuations. The PINNs architecture addresses this coupled system through a comprehensive loss function that incorporates acoustic wave propagation, flame dynamics, mean flow effects, and boundary conditions, as expressed in Eq.~\eqref{eq:loss_function}:

\begin{equation}
\mathcal{L}_{total} = \mathcal{L}_{acoustic} + \mathcal{L}_{flame} + \mathcal{L}_{coupling} + \mathcal{L}_{data}
\label{eq:loss_function}
\end{equation}

\subsubsection{PINNs Implementation for Thermoacoustic Systems}

Modern PINNs implementations for thermoacoustic systems integrate experimental measurements with governing physical equations to reconstruct the complete acoustic field. Mariappan et al.~\cite{mariappan2024learning} demonstrated a PINNs approach for bluff body anchored flame combustors, where acoustic pressure measurements at sparse locations and total flame heat release data served as inputs. Their coupled parameterized model combined acoustic equations with a van der Pol oscillator to capture vortex shedding dynamics, as shown in Eq.~\eqref{eq:van_der_pol}:

\begin{equation}
\frac{d^2q_{pmt}}{dt^2} - \mu(1-q_{pmt}^2)\frac{dq_{pmt}}{dt} + \omega_v^2q_{pmt} - \gamma u(x_r,t) = 0
\label{eq:van_der_pol}
\end{equation}

\begin{figure}[htbp]
    \centering
    \includegraphics[width=0.9\textwidth]{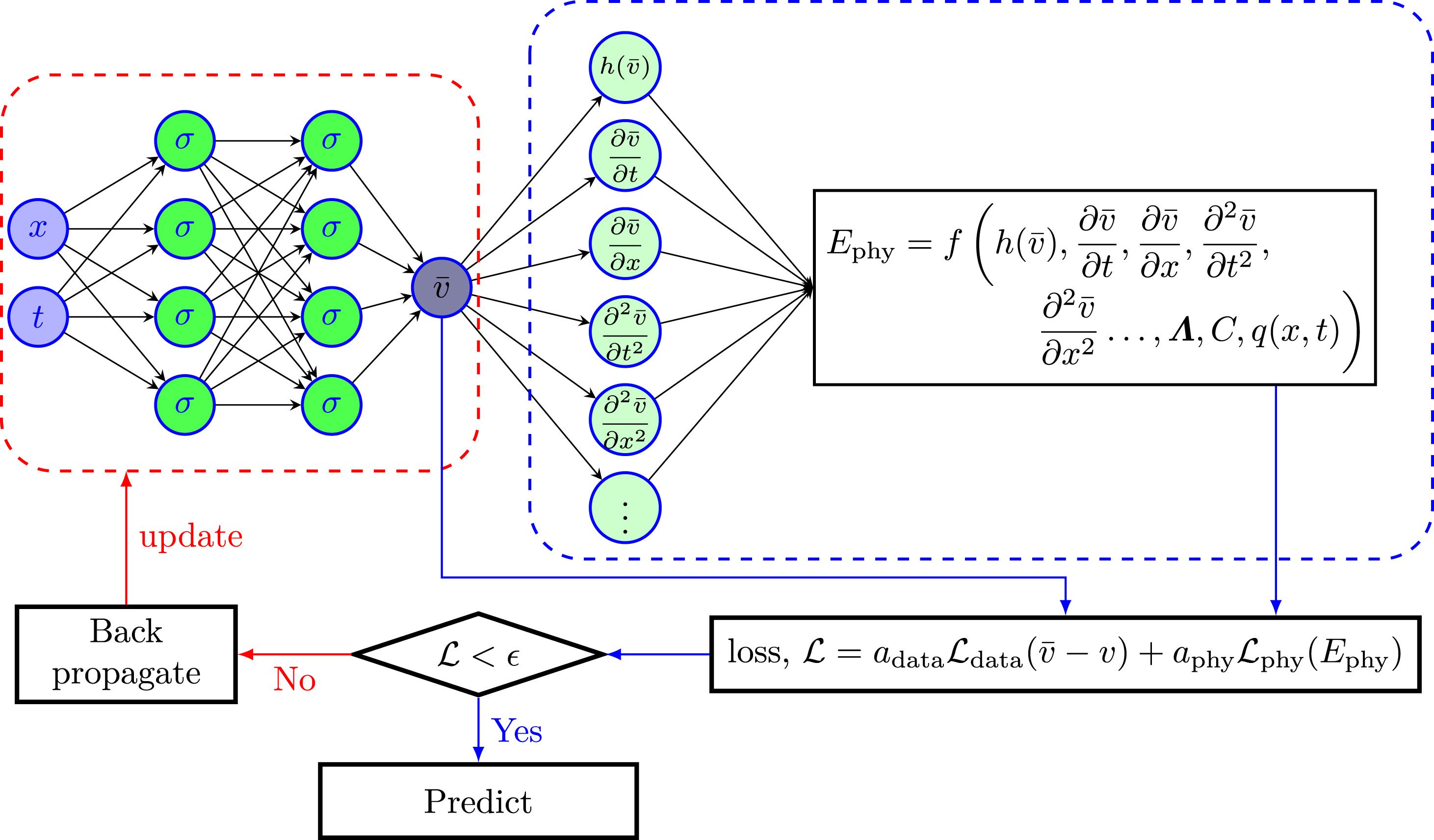}
    \caption{Schematic of PINNs for inverse problems in thermoacoustic systems. The diagram shows a deep neural network (left) with inputs of space and time coordinates, providing outputs that are constrained by both physical equations (right) and experimental data. Adapted from Mariappan et al. \cite{mariappan2024learning}.}
    \label{fig:pinn_schematic}
\end{figure}

This formulation in Eq.~\eqref{eq:van_der_pol} addresses several practical challenges in thermoacoustic modeling, including unknown initial/boundary conditions, absence of acoustic velocity measurements, and variation of acoustic pressure amplitude over long time scales due to turbulent flame behavior \cite{lieuwen2003, juniper2023}. Figure~\ref{fig:pinn_schematic} illustrates the schematic of the PINNs framework applied to thermoacoustic problems, showing how both physical equations and experimental data constrain the neural network training process. To overcome implementation challenges, Mariappan et al. \cite{mariappan2024learning} have developed innovative enhancements to the PINNs training procedure \cite{mariappan2024learning}. These include systematic adjustment of the relative contribution of various loss terms, inclusion of loss terms associated with acoustic velocity to ensure uniqueness, and time series segmentation to handle temporal variations in amplitude \cite{mariappan2024learning}.

\subsubsection{Experimental Validation and Performance Assessment}

The validation of PINNs-based stability predictions involves rigorous comparison with experimental measurements and classical stability analysis methods \cite{dawson2014, singh2021}. The accuracy of these methods can be quantified through comprehensive error metrics as shown in Eq.~\eqref{eq:error_metric}:

\begin{equation}
\epsilon = \sqrt{\frac{1}{N}\sum_{i=1}^N (f_{pred,i} - f_{exp,i})^2}
\label{eq:error_metric}
\end{equation}

\begin{figure}[htbp]
    \centering
    \includegraphics[width=0.9\textwidth]{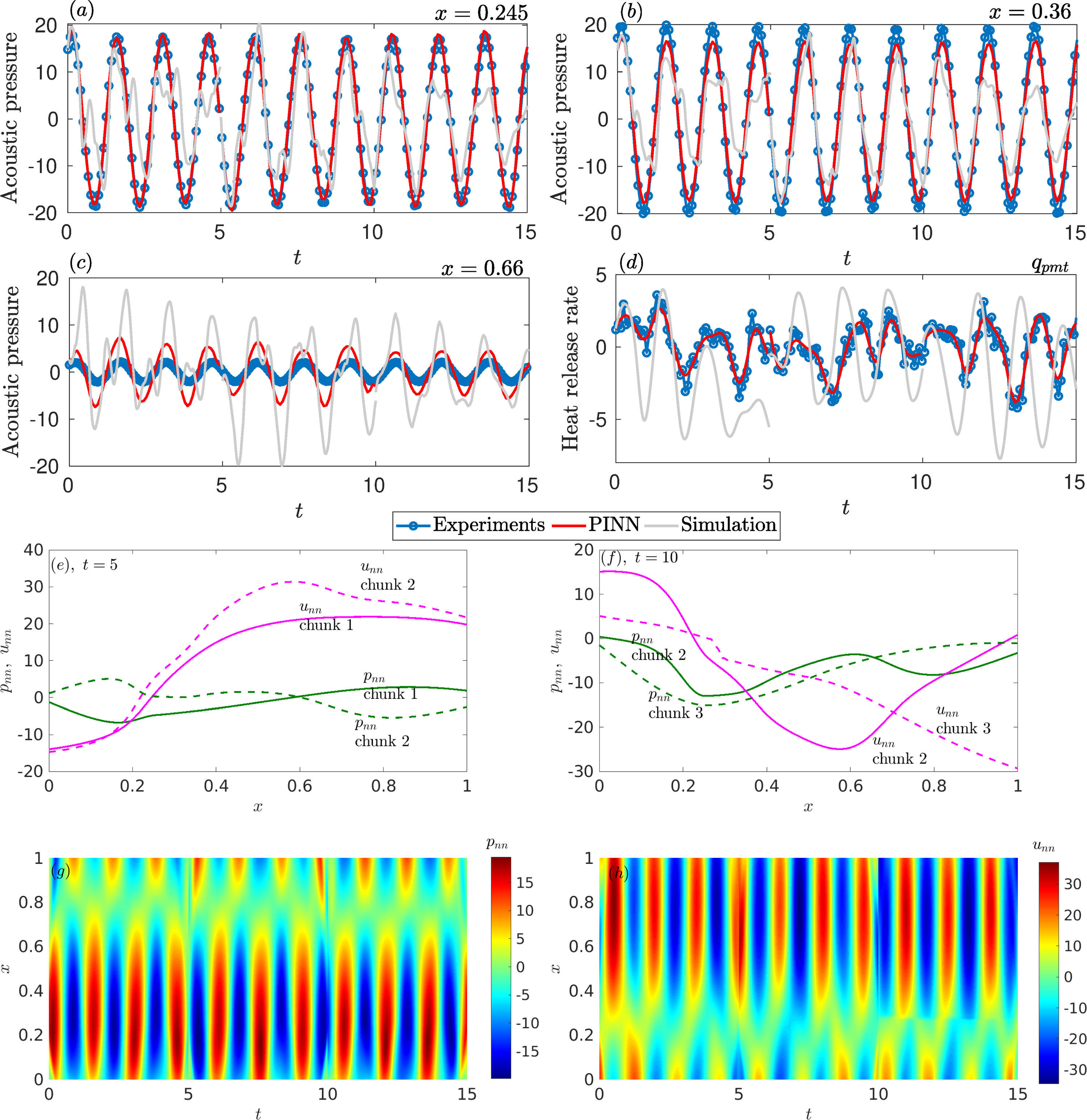}
    \caption{PINNs solution from experimental data showing (a-d) comparison of acoustic pressure and heat release rate between experiments, PINNs prediction, and simulation; (e-f) spatial distribution of pressure and velocity at chunk intersections; (g-h) complete spatiotemporal evolution of acoustic pressure and velocity fields. Adapted from Mariappan et al. \cite{mariappan2024learning}.}
    \label{fig:pinn_results}
\end{figure}

Mariappan et al.~\cite{mariappan2024learning} demonstrated that PINN-based approaches can accurately reconstruct acoustic fields and estimate model parameters in practical combustors, even when limited experimental measurements are available, as shown in Fig.~\ref{fig:pinn_results}. This capability is particularly valuable for predicting the spatial distribution of acoustic pressure and velocity—critical parameters for structural design and thermal protection systems \cite{culick1988, richecoeur2008}.

The error metric defined in Eq.~\eqref{eq:error_metric} provides a quantitative assessment of the prediction fidelity across different operating regimes. As illustrated in Fig.~\ref{fig:pinn_results}, PINNs can successfully reconstruct complete acoustic fields from sparse measurement data, offering detailed spatial and temporal profiles of pressure and velocity. These reconstructions enable insights that are difficult to obtain through direct measurements alone, enhancing the analysis and design of combustion systems subject to acoustic instabilities.

\begin{figure}[htbp]
    \centering
    \includegraphics[width=0.8\textwidth]{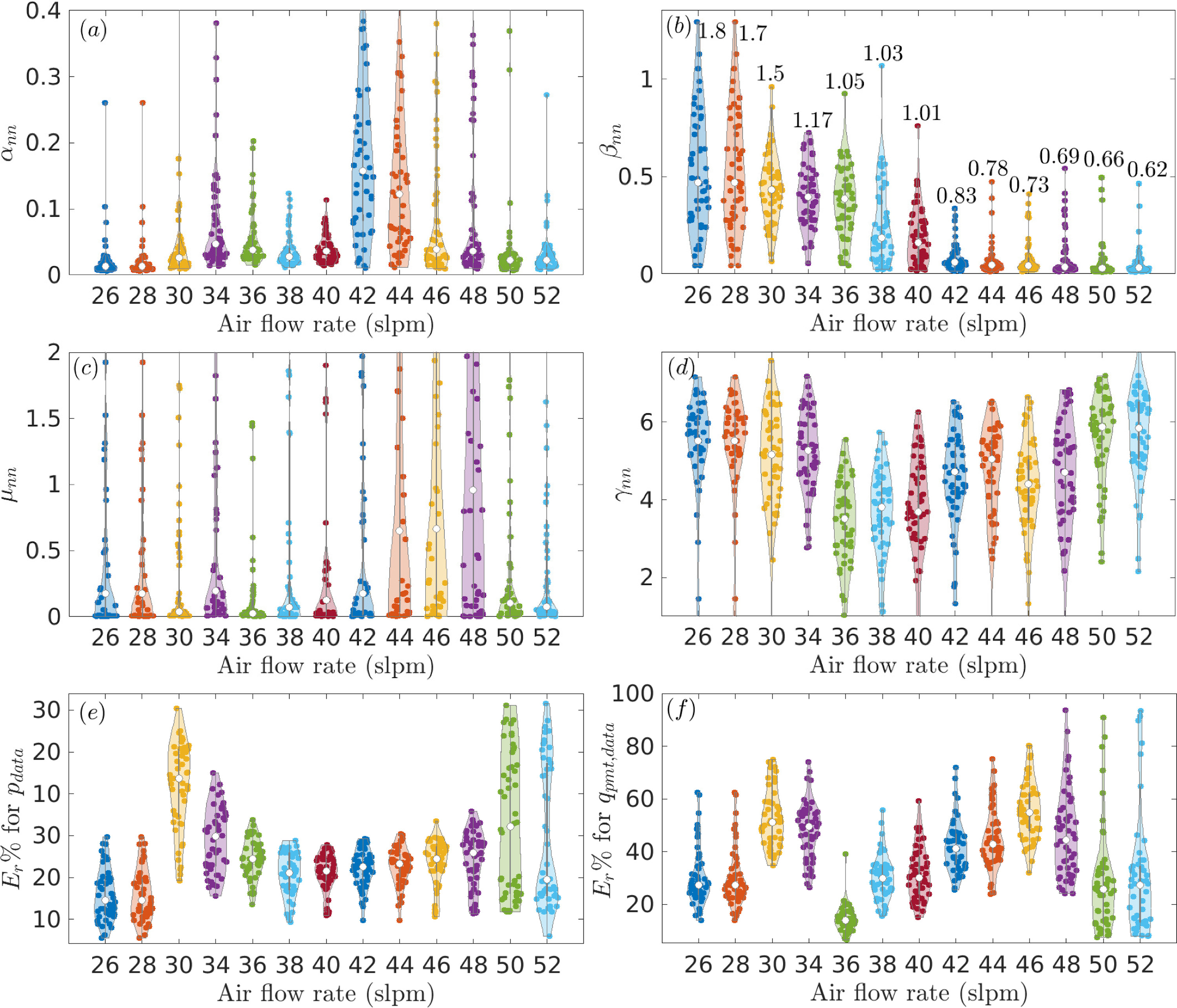}
    \caption{Probability density functions of the LOM parameters and train errors predicted by PINNs across different airflow rates, showing (a-d) estimated model parameters for acoustic damping, heat release coupling, nonlinearity, and acoustic velocity coupling; (e-f) training errors for pressure and heat release data. Adapted from Mariappan et al. \cite{mariappan2024learning}.}
    \label{fig:parameter_distribution}
\end{figure}

Beyond field reconstruction, PINNs facilitate parameter identification for low-order models, offering valuable insights into the physical mechanisms driving thermoacoustic instabilities. As shown in Fig.~\ref{fig:parameter_distribution}, the distribution of estimated model parameters across different operating conditions highlights variations in key factors such as acoustic damping, heat release coupling, nonlinearity, and acoustic velocity coupling. Analyzing these parameter trends enables a deeper understanding of the transitions between stable and unstable regimes, supporting the development of more effective control strategies to mitigate combustion instabilities.

\subsubsection{Emerging Research Directions}

Ongoing research in PINNs for combustion instabilities is advancing toward uncertainty quantification and comprehensive multi-physics modeling \cite{karniadakis2021physics, li2023}. Key developments include real-time prediction algorithms that enable adaptive control \cite{zhang2022}, robust model-order reduction techniques to enhance computational efficiency \cite{jagtap2020}, and the integration of Bayesian inference methods for quantifying uncertainty \cite{garita2021, juniper2022}. Additionally, PINNs are increasingly being applied to combustion systems using novel fuels such as hydrogen and ammonia \cite{valera-medina2019}, addressing challenges unique to these alternative energy carriers.

The fusion of physics-based modeling and machine learning within the PINNs framework represents a transformative shift in the analysis and control of combustion instabilities \cite{mariappan2024learning}. This approach delivers high-fidelity predictions with computational efficiency suitable for real-time applications. 

\subsection{Solving Stiff Reaction Systems}

Stiff reaction systems are commonly encountered in chemical and combustion processes. These systems are typically governed by ordinary differential equations (ODEs) of the form:

\begin{equation}
\frac{d\mathbf{Y}}{dt} = \mathbf{R}(\mathbf{Y}; \boldsymbol{\kappa})
\end{equation}

Here, $\mathbf{Y}$ denotes the vector of species concentrations, $\mathbf{R}$ represents the reaction rate terms, and $\boldsymbol{\kappa}$ includes the reaction rate constants. These systems often involve vastly different time scales, leading to stiffness and significant computational challenges for conventional numerical solvers \cite{zhang2024crk,sun2023physics,wu2023application}.

PINNs have emerged as a promising alternative for solving stiff reaction systems. PINNs are trained to satisfy the governing equations by embedding their residuals within the loss function. A neural network $\mathbf{N}(t; \boldsymbol{\theta})$, parameterized by $\boldsymbol{\theta}$, approximates the solution $\mathbf{Y}(t)$. The loss function $\mathcal{L}$ typically includes terms enforcing both the governing equations and data constraints:

\begin{equation}
\mathcal{L} = \sum_{i} \left\| \frac{d\mathbf{N}(t_i; \boldsymbol{\theta})}{dt} - \mathbf{R}(\mathbf{N}(t_i; \boldsymbol{\theta}); \boldsymbol{\kappa}) \right\|^2
\end{equation}

Despite their flexibility, standard PINNs face difficulties when applied to stiff systems due to multiscale dynamics, steep gradients, and the potential for vanishing gradients during training. To overcome these limitations, several approaches have been proposed in recent studies.

\vspace{1em}
\subsubsection{CRK-PINNs for Stiff Reaction Systems}

CRK-PINNs introduce logarithmic normalization of species mole fractions, temperature, and pressure to address the multiscale characteristics of stiff reaction systems. In this approach, species mole fractions are normalized as:

\begin{equation}
X'_k = \frac{\ln(X_k) - \mu[\ln(X_k)]}{\sigma[\ln(X_k)]}
\end{equation}

where $\mu$ and $\sigma$ are the mean and standard deviation, respectively, of the logarithmic mole fractions \cite{zhang2024crk}. This normalization reduces ill-conditioning and facilitates faster convergence during training. CRK-PINNs also integrate physical constraints, including enthalpy conservation, element conservation, and mole fraction conservation, directly into their loss function:

\begin{equation}
\mathcal{L} = \lambda_1 \mathcal{L}_{ODEs} + \lambda_2 \mathcal{L}_{Enthalpy} + \lambda_3 \mathcal{L}_{Elements} + \lambda_4 \mathcal{L}_{Mole} + \lambda_5 \mathcal{L}_{Data}
\end{equation}

\begin{figure}[h]
    \centering
    \includegraphics[width=0.8\textwidth]{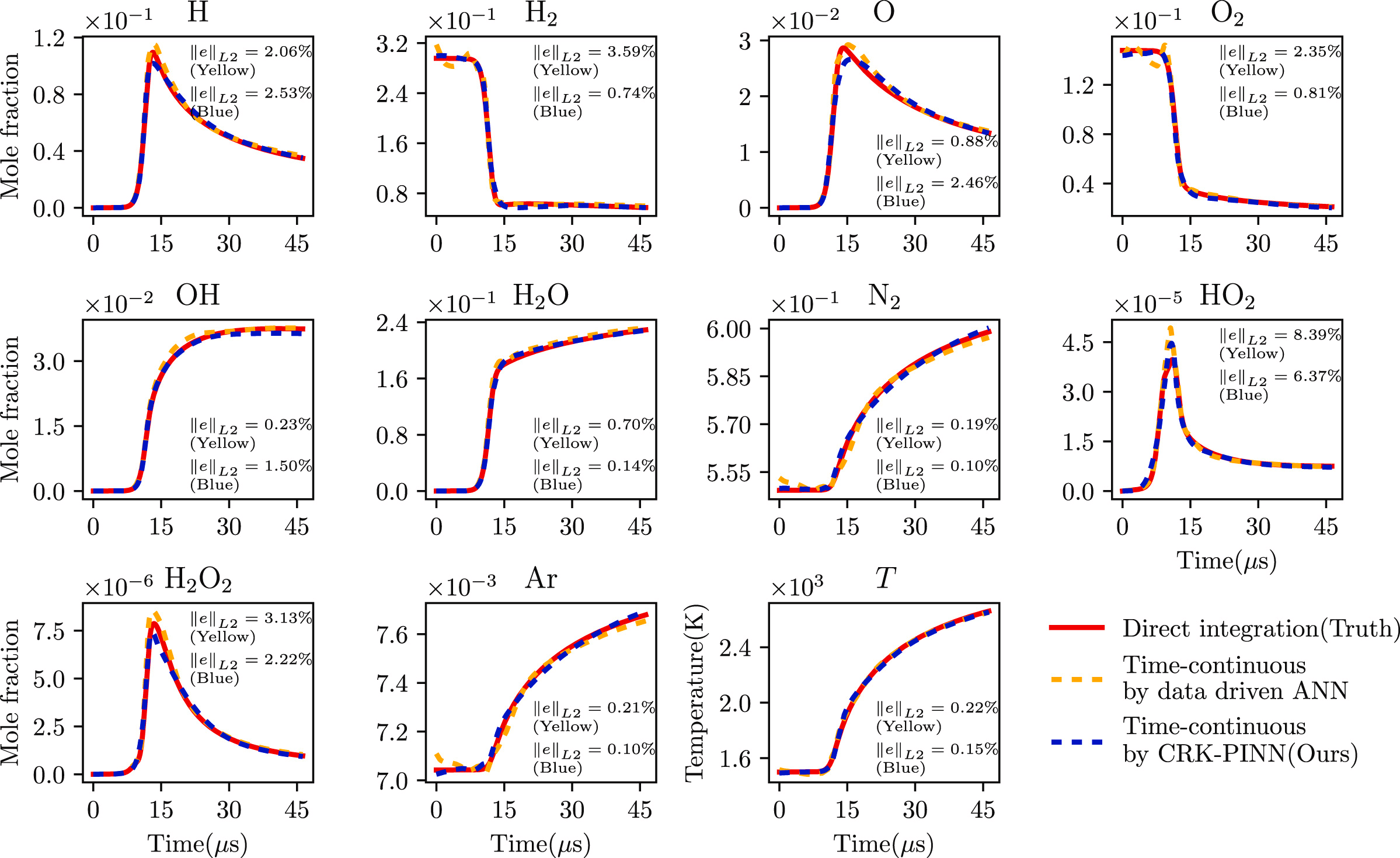}
    \caption{Time-continuous predictions of the autoignition process via CRK-PINNs and data-driven ANN. Adapted from Zhang et al. (2024) \cite{zhang2024crk}.}
    \label{fig:crkpinn_framework}
\end{figure}

The framework of CRK-PINNs is illustrated in Fig.~\ref{fig:crkpinn_framework}, demonstrating their ability to handle stiff reaction kinetics with reduced dependency on training data and improved accuracy \cite{zhang2024crk}.

\vspace{1em}
\subsubsection{DPINNs for Convection-Dominated and Stiff CDR Problems}

DPINNs (Derivative-constrained PINNs) further enhance the accuracy and stability of PINNs when applied to convection-dominated and stiff convection-diffusion-reaction (CDR) problems \cite{hoshisashi2024physicsinformed}. They introduce a derivative constraint loss term, incorporating the spatial first derivative of the residuals of the governing equations into the loss function:

\begin{equation}
\mathcal{L}_{fx} = \frac{1}{|X_f|} \sum_{X \in X_f} \left\| \frac{\partial}{\partial x} \left( \frac{\partial u_\theta}{\partial t} + v \frac{\partial u_\theta}{\partial x} - D \frac{\partial^2 u_\theta}{\partial x^2} - R \right) \right\|^2
\end{equation}

This approach smooths the PDE solution and suppresses non-physical oscillations caused by sharp spatial gradients \cite{sun2023physics}. DPINNs also employ deep function family construction, where the input is concatenated with outputs of hidden layers, enhancing the networks ability to approximate complex nonlinear functions. The final DPINN loss function is:

\begin{equation}
\mathcal{L} = \omega_s \mathcal{L}_s + \omega_{fx} \mathcal{L}_{fx} + \omega_v \mathcal{L}_v
\end{equation}

\begin{figure}[H]
\centering
% Subfigure (a)
\begin{subfigure}{0.32\textwidth}
    \centering
    \includegraphics[width=\textwidth]{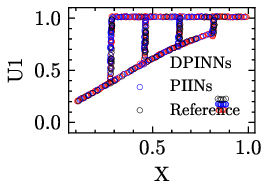}
    \subcaption{U1}
    \label{fig:6A}
\end{subfigure}
\hfill
% Subfigure (b)
\begin{subfigure}{0.32\textwidth}
    \centering
    \includegraphics[width=\textwidth]{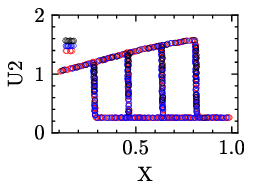}
    \subcaption{U2}
    \label{fig:6B}
\end{subfigure}
\hfill
% Subfigure (c)
\begin{subfigure}{0.32\textwidth}
    \centering
    \includegraphics[width=\textwidth]{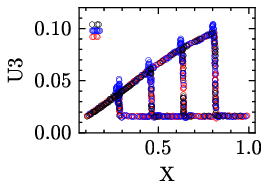}
    \subcaption{U3}
    \label{fig:6C}
\end{subfigure}
\caption{Comparison of DPINNs and PINNs solutions against the reference at four time levels. Adapted from Sun et al. \cite{sun2023physics}.}
\label{fig:dpinn_comparison}
\end{figure}

Figure~\ref{fig:dpinn_comparison} shows a comparison between PINNs and DPINNs solutions against the reference at four different time levels, demonstrating the superior ability of DPINNs to capture steep gradients and reduce oscillations \cite{sun2023physics}.

\begin{figure}[h]
    \centering
    \includegraphics[width=0.7\textwidth]{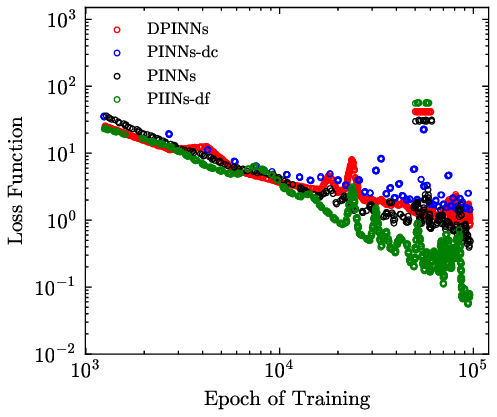}
    \caption{Comparison of loss functions for different PINN-based methods, including DPINN. Adapted from Sun et al. \cite{sun2023physics}.}
    \label{fig:dpinn_loss}
\end{figure}

Further analysis of the loss functions for different PINNs approaches, including PINNs with derivative constraint (PINN-dc), PINNs with deep function family construction (PINN-df), and DPINN, indicates the superior convergence and stability of DPINNs. This is illustrated in Fig.~\ref{fig:dpinn_loss} \cite{sun2023physics}.

\vspace{1em}
\subsubsection{Quasi-Steady-State Assumption to Reduce Stiffness}

The quasi-steady-state assumption (QSSA) can be applied to further reduce the stiffness of reaction systems. This approach separates species into fast and slow groups, assuming that the fast species reach equilibrium rapidly. The QSSA model is expressed as:

\begin{equation}
\begin{aligned}
\frac{d\mathbf{Y}_s}{dt} &= \mathbf{R}_s(\mathbf{Y}_s, \mathbf{Y}_f; \boldsymbol{\kappa}) \\
0 &= \mathbf{R}_f(\mathbf{Y}_s, \mathbf{Y}_f; \boldsymbol{\kappa})
\end{aligned}
\end{equation}

Applying QSSA simplifies the system, reduces computational stiffness, and improves the efficiency of PINNs training \cite{zhang2024crk, wu2023application}.

\vspace{1em}
\subsubsection{Performance and Advantages of CRK-PINNs and DPINNs}

PINN-based methods, including CRK-PINNs and DPINNs, demonstrate significant computational advantages over conventional numerical solvers. CRK-PINNs achieved speedups of 6.0 to 14.6 times for chemical source term integration and 2.3 to 4.9 times for reactive flow simulations compared to traditional methods \cite{zhang2024crk}. DPINNs eliminate spurious oscillations in convection-dominated flows and accurately capture steep gradients in autocatalytic reaction systems, resulting in enhanced stability and convergence efficiency \cite{sun2023physics}.

\subsection{Optimization and Control of Combustion Systems}

The integration of PINNs with advanced control strategies represents a significant advancement in the optimization and control of combustion systems. Recent studies have demonstrated that PINNs enhance system efficiency, stability, and environmental performance \cite{zhu2024physics, taassob2024pinn}. Their ability to provide accurate predictions and reduce computational costs makes them particularly effective for model predictive control (MPC) applications. 

\subsubsection{PINN-Enhanced Model Predictive Control}

PINN-enhanced MPC has emerged as a robust strategy for controlling combustion systems under dynamic and uncertain operating conditions. The fundamental MPC optimization problem is formulated as \cite {paredes2024mpc}
\begin{equation}
\min_{u(t)} \int_{t_0}^{t_f} L(x(t), u(t))dt + V_f(x(t_f))
\end{equation}
subject to the system dynamics governed by the PINN-based model,
\begin{equation}
\dot{x}(t) = f_{\text{PINN}}(x(t), u(t))
\end{equation}
where \(L\) is the stage cost, \(V_f\) is the terminal cost, \(x(t)\) is the system state, and \(u(t)\) is the control input. The function \(f_{\text{PINN}}\) encapsulates the physics-informed dynamic model, enabling rapid and accurate state predictions \cite{zhu2024physics, taassob2024pinn}.

In the receding horizon framework, the optimal control input is determined by solving
\begin{equation}
u^*(t) = \arg\min_{u(t)} \sum_{k=0}^{N_p} \|x_k - x_{\text{ref}}\|_Q + \|u_k\|_R
\end{equation}

where \(N_p\) is the prediction horizon, \(Q\) and \(R\) are weighting matrices, and \(x_{\text{ref}}\) is the reference trajectory. The integration of PINNs enhances the MPC formulation by ensuring physical consistency in state prediction, as demonstrated in the work of Zhu et al. \cite{zhu2024physics}, where monotonic physical relationships are embedded within the PINNs architecture for improved generalization.

\subsubsection{Real-time Implementation and Adaptation}

Real-time control of combustion systems requires computationally efficient models that maintain accuracy across a wide range of operating conditions. PINN-based architectures support real-time evaluation of system dynamics by employing neural networks parameterized by \(\theta\):
\begin{equation}
\hat{x}_{k+1} = \mathcal{N}_{\theta}(x_k, u_k)
\end{equation}

The total loss function used in the training process typically combines multiple components, ensuring consistency with available data, physical laws, and operational constraints:
\begin{equation}
\mathcal{L}_{\text{total}} = \mathcal{L}_{\text{data}} + \mathcal{L}_{\text{physics}} + \mathcal{L}_{\text{constraint}}
\end{equation}

\begin{figure}[htbp]
    \centering
    \includegraphics[width=0.8\textwidth]{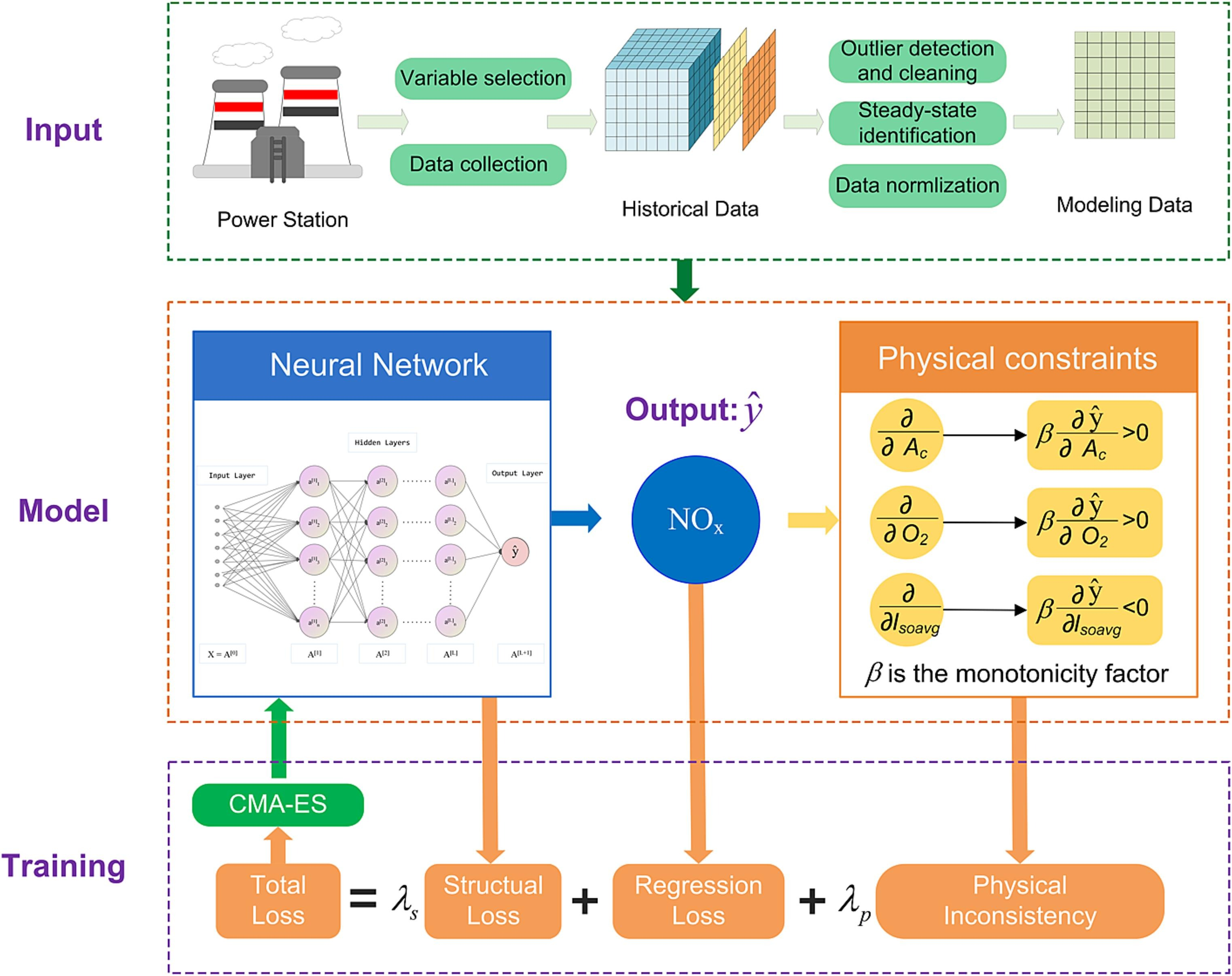}
    \caption{PINN-based NO$_x$ emission prediction framework from Zhu et al. \cite{zhu2024physics}.}
    \label{fig:pinn_framework_zhu}
\end{figure}

As demonstrated by Taassob et al. \cite{taassob2024pinn}, the DeepONet-PINNs framework extends traditional PINNs by effectively handling complex parametric dependencies, such as variations in inlet velocities and jet recess lengths in turbulent combustion systems. Similarly, Zhu et al. \cite{zhu2024physics} introduced a comprehensive PINN-based model for real-time NO$_x$ emission prediction (see Fig.~\ref{fig:pinn_framework_zhu}), underscoring the architecture’s suitability for real-time deployment.

Adaptation to changing operational conditions can be achieved through online learning techniques, where network parameters are updated dynamically to maintain model accuracy:
\begin{equation}
\frac{d\theta}{dt} = -\eta \nabla_{\theta} \mathcal{L}_{\text{total}}
\end{equation}

This adaptive learning mechanism enables the system to respond effectively to variations in combustion dynamics and operational demands.

\subsubsection{Efficiency Optimization}

\begin{figure}[htbp]
    \centering
    \includegraphics[width=0.5\textwidth]{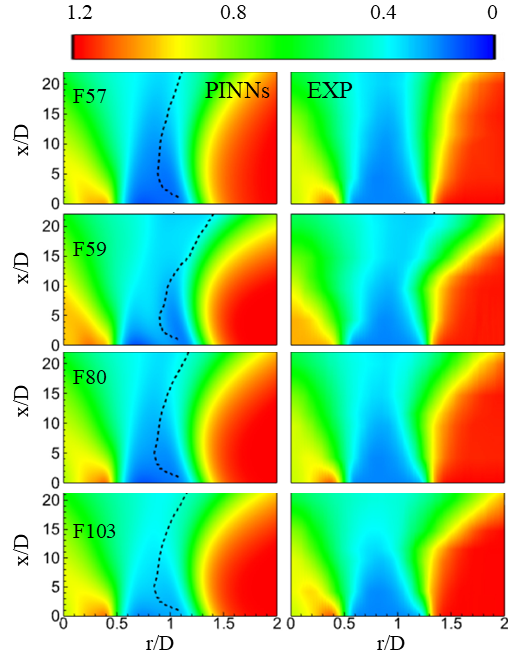}
    \caption{Comparisons of density contours for Flames 57, 59, 80, and 103 \cite{taassob2024pinn}.}
    \label{fig:density_contours}
\end{figure}

\begin{figure}[htbp]
    \centering
    \includegraphics[width=0.5\textwidth]{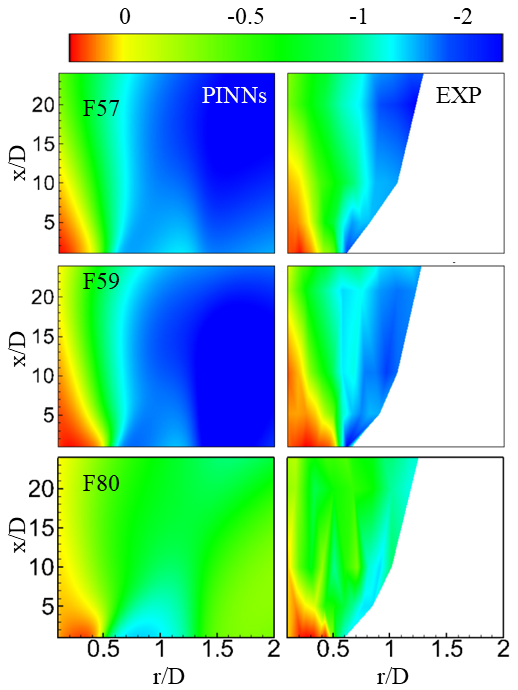}
    \caption{Comparisons of radial velocity contours for Flames 57, 59, and 80 \cite{taassob2024pinn}.}
    \label{fig:radial_velocity_contours}
\end{figure}

Optimizing combustion efficiency is a key objective in modern energy systems. The efficiency metric can be expressed as
\begin{equation}
\eta_{\text{combustion}} = \frac{\dot{Q}_{\text{out}}}{\dot{m}_{\text{fuel}} \cdot LHV}
\end{equation}
where \(\dot{Q}_{\text{out}}\) is the heat release rate, \(\dot{m}_{\text{fuel}}\) is the fuel mass flow rate, and \(LHV\) is the lower heating value of the fuel. Zhu et al. \cite{zhu2024physics} demonstrated how PINNs can be utilized to enforce physical laws and constraints, ensuring that the optimization respects operational boundaries and maintains emissions within acceptable limits. The optimization problem can be formalized as
\begin{equation}
\max_{u(t)} \eta_{\text{combustion}}(u(t))
\end{equation}
subject to
\begin{equation}
g(x(t), u(t)) \le 0
\end{equation}
where \(g\) represents constraint functions related to emissions, stability, and safety. Taassob et al. \cite{taassob2024pinn} demonstrated the efficacy of their PINN-DeepONet framework through the accurate prediction of density and velocity fields for multiple flame configurations. The density contours for different flames, including Flames 57, 59, 80, and 103 (Fig.~\ref{fig:density_contours}), alongside the radial velocity contours (Fig.~\ref{fig:radial_velocity_contours}), illustrate the predictive accuracy and generalization capabilities of their approach.

\subsubsection{Sustainability and Emissions Control}

Reducing pollutant emissions while maintaining high efficiency is a central challenge in combustion system design. PINNs enable multi-objective optimization that balances efficiency and emissions. The optimization problem can be expressed as
\begin{equation}
\min_{u(t)} \left\{ -\eta_{\text{combustion}}(u(t)), E_{\text{NOx}}(u(t)), E_{\text{CO}}(u(t)) \right\}
\end{equation}
Zhu et al. \cite{zhu2024physics} embedded monotonic relationships within their PINNs model to predict NO$_x$ emissions from coal-fired boilers, demonstrating improved prediction accuracy and robustness compared to traditional machine learning methods. The optimal control strategy must satisfy
\begin{equation}
u^*(t) = \arg\min_{u(t)} J(x(t), u(t))
\end{equation}
subject to emission constraints
\begin{equation}
E_i(t) \le E_{i, \text{max}}, \quad i \in \{ \text{NOx}, \text{CO}, \text{PM} \}
\end{equation}
Taassob et al. \cite{taassob2024pinn} further emphasized the use of PINN-DeepONet for recovering species reaction rates and turbulent closures, ensuring accurate representation of pollutant formation in complex flame configurations.

\subsubsection{System Design Optimization}

The application of PINNs in system design extends beyond operational control to include optimization of geometry and operating parameters. The design optimization problem is formulated as
\begin{equation}
\min_{p} \{ f_1(p), f_2(p), ..., f_m(p) \}
\end{equation}
where \(p\) are design parameters and \(f_i\) are objective functions for efficiency, emissions, and stability. PINNs act as surrogate models, enabling rapid evaluation of design alternatives:
\begin{equation}
\hat{y} = \mathcal{N}_{\theta}(p)
\end{equation}
Taassob et al. \cite{taassob2024pinn} showcased the capability of PINNs in representing complex dependencies between input parameters and output responses, facilitating efficient exploration of the design space.

\subsubsection{Robust Control and Uncertainty Management}

Robust control requires accounting for uncertainties in system dynamics and operating conditions. The robust optimization problem can be formulated as
\begin{equation}
\min_{u(t)} \max_{w \in \mathcal{W}} J(x(t), u(t), w)
\end{equation}
where \(w\) represents uncertain parameters and \(\mathcal{W}\) is the uncertainty set. Taassob et al. \cite{taassob2024pinn} demonstrated that PINN-DeepONet architectures can effectively handle variations in boundary conditions and operating parameters, ensuring stability and robustness. Lyapunov-based constraints can be employed to guarantee system stability:
\begin{equation}
\frac{dV}{dt} + \alpha V \le 0
\end{equation}
where \(V\) is a Lyapunov function and \(\alpha\) ensures exponential stability.

\subsubsection{Advanced Control Architectures}

\begin{figure}[htbp]
    \centering
    \includegraphics[width=0.6\textwidth]{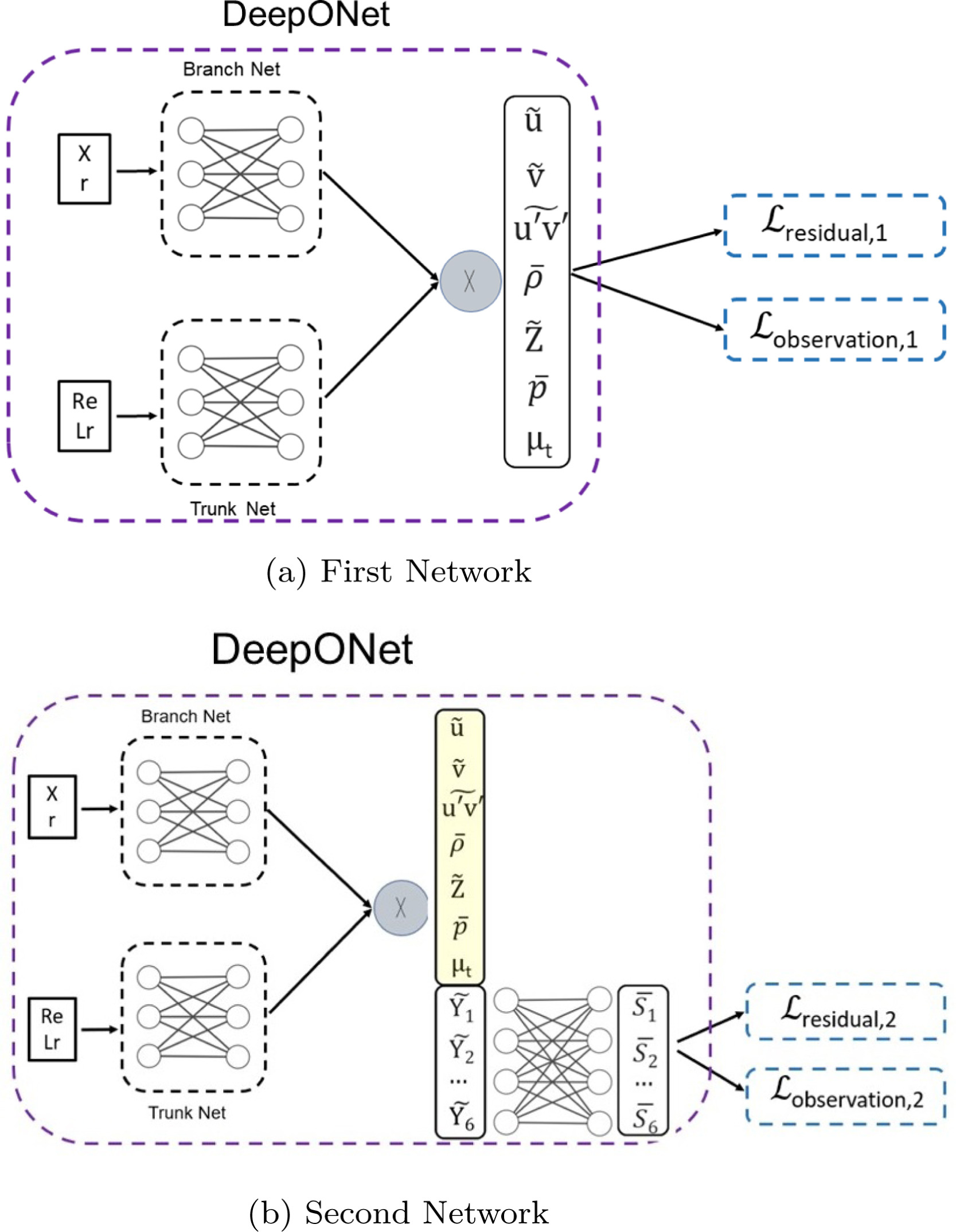}
    \caption{Schematic of the two PINN-DeepONet networks as presented by Taassob et al. \cite{taassob2024pinn}.}
    \label{fig:pinn_deeponet_schematic}
\end{figure}

The hierarchical integration of PINNs within advanced control frameworks enables comprehensive optimization across multiple control layers. Each layer addresses a specific control objective:
\begin{equation}
u(t) = \{ u_1(t), u_2(t), ..., u_L(t) \}
\end{equation}
The dynamic behavior of each layer is described by
\begin{equation}
\dot{x}_i = f_i(x_i, u_i, v_i)
\end{equation}
where \(v_i\) are coordination variables between layers. Taassob et al. \cite{taassob2024pinn} illustrated how their two-tier PINN-DeepONet architecture (Fig.~\ref{fig:pinn_deeponet_schematic}) can be modularly deployed, allowing flexible and scalable control strategies that incorporate both data-driven insights and physics-based constraints.

The continued development and integration of PINN-based optimization and control strategies promise significant advancements in combustion efficiency, stability, and environmental performance \cite{toscano2025pinns}. By leveraging physics-based modeling and advanced neural network architectures, PINNs offer a pathway to next-generation combustion systems capable of meeting stringent regulatory requirements while maintaining robust and reliable operation.

% Figures (add these to your LaTeX document at the end)

\section{Comparison with Traditional Combustion Modeling Techniques}

The emergence of PINNs has initiated a reevaluation of combustion modeling methodologies, offering novel approaches to simulate complex reactive flows. This section presents a comprehensive comparison between PINNs-based models and traditional techniques, including CFD employing FVM and Finite Element Methods (FEM), as widely used in combustion modeling \cite{song2024physics, zhang2024crk, mariappan2024learning}.

\subsection{Fundamental Differences in Approach}

Traditional CFD techniques numerically solve conservation equations by discretizing space and time. For example, FVM subdivides the domain into control volumes and integrates conservation laws:
\begin{equation}
\frac{\partial}{\partial t}\int_V \rho \phi \, dV + \oint_A \rho \phi \, \mathbf{v} \cdot d\mathbf{A} = \oint_A \Gamma \nabla \phi \cdot d\mathbf{A} + \int_V S_{\phi} \, dV
\end{equation}
where \( \phi \) is a generic scalar quantity, \( \Gamma \) the diffusion coefficient, and \( S_{\phi} \) the source term.

By contrast, PINNs utilize neural networks to approximate solutions by minimizing a composite loss function. This formulation enforces consistency with both data and the underlying physical principles:
\begin{equation}
\mathcal{L}_{\text{total}} = \mathcal{L}_{\text{data}} + \lambda \mathcal{L}_{\text{physics}}
\end{equation}
Unlike mesh-based methods, PINNs operate on scattered data points and can handle complex geometries without explicit domain discretization \cite{raissi2019physics, nasiri2022reduced}.

\subsection{Computational Efficiency and Resource Requirements}

High-fidelity CFD combustion simulations often require substantial computational resources due to fine mesh resolution and the need for small time steps:
\begin{equation}
t_{\text{comp}} \propto N_x \times N_y \times N_z \times N_t
\end{equation}

PINNs eliminate the meshing step and, after the training phase, allow for rapid evaluations via forward passes through the trained network:
\begin{equation}
t_{\text{eval}} \propto N_{\text{layers}} \times N_{\text{neurons}}
\end{equation}

Nevertheless, training PINNs—especially for stiff chemical systems and multi-physics problems—can be computationally demanding. Recent advances, such as Reduced-PINNs employing integral formulations, have demonstrated improved efficiency in handling these challenges \cite{nasiri2022reduced}.

\subsection{Accuracy and Solution Quality}

Traditional methods, such as FVM and FEM, offer well-established error estimates and convergence properties. The discretization error follows known relations:
\begin{equation}
\varepsilon \propto (\Delta x)^n + (\Delta t)^m
\end{equation}
where \( n \) and \( m \) are the orders of spatial and temporal schemes, respectively.

PINNs provide continuous solution representations:
\begin{equation}
u(x, t) = \mathcal{N}_{\theta}(x, t)
\end{equation}
This smooth approximation enables automatic differentiation and efficient gradient computation. However, accuracy is highly dependent on network architecture, training data quality, and optimization algorithms \cite{de2022error}. Zhang et al. \cite{zhang2024crk} introduced CRK-PINN, integrating conservation laws explicitly into the loss function to improve the physical accuracy of predictions and reduce non-physical behavior in combustion simulations.

\subsection{Handling of Complex Geometries}

Mesh generation for complex geometries in traditional CFD requires significant effort and expertise. The mesh quality, characterized by skewness, aspect ratio, and orthogonality, directly impacts numerical accuracy.

PINNs inherently avoid mesh dependency by defining boundary conditions as constraints in the loss function as follows.
\begin{equation}
\mathcal{L}_{\text{boundary}} = \|u(x_b,t) - g(x_b,t)\|^2
\end{equation}
Mariappan et al. \cite{mariappan2024learning} successfully applied PINNs to reconstruct acoustic fields in complex combustor geometries without relying on structured meshes, demonstrating their potential in thermoacoustic modeling.

\subsection{Treatment of Multi-Scale Phenomena}

Combustion processes span multiple spatial and temporal scales. Traditional CFD addresses these scales through adaptive mesh refinement (AMR) and multi-grid methods:
\begin{equation}
\Delta x_{\text{local}} = f(\text{gradient}, \text{curvature}, \text{reaction rates})
\end{equation}

PINNs can capture multi-scale features by designing network architectures and loss functions that account for different physical processes:
\begin{equation}
\mathcal{L}_{\text{multi-scale}} = \sum_i w_i\mathcal{L}_i
\end{equation}
Song et al. \cite{song2024physics} and Ji et al. \cite{ji2021stiff} demonstrated PINNs capability to resolve multi-scale combustion processes through physics-informed loss balancing and adaptive weighting strategies.

\subsection{Hybrid Approaches and Complementary Integration}

Recognizing the complementary strengths of traditional CFD and PINNs, hybrid frameworks are increasingly explored. One such integration leverages PINNs for modeling subgrid-scale (SGS) phenomena in Large Eddy Simulation (LES):
\begin{equation}
\tau_{\text{sgs}} = \mathcal{N}_{\theta}(\nabla u, Re, \Delta)
\end{equation}
Wu et al. \cite{wu2023application} demonstrated hybrid models using PINNs for chemical source terms while employing conventional CFD solvers for flow field calculations:
\begin{equation}
\dot{\omega}_i = \mathcal{N}_{\theta}(Y_i, T, P)
\end{equation}

These approaches reduce computational overhead in stiff chemistry simulations and enable high-fidelity predictions at reduced costs.

\subsection{Implementation and Practical Considerations}

Conventional CFD software—such as OpenFOAM, ANSYS Fluent, and COMSOL Multiphysics—has undergone extensive validation across decades of development. These platforms offer robust solvers, comprehensive documentation, and support from well-established user communities.

By comparison, implementing PINNs requires a synergistic understanding of machine learning techniques and combustion system dynamics. The design of an effective architecture involves selecting the number of layers, neurons, and activation functions:
\begin{equation}
\mathcal{A} = f(N_{\text{layers}}, N_{\text{neurons}}, \text{activation functions})
\end{equation}

Equally important is the formulation of a stable and convergent training strategy, which minimizes the composite loss function:
\begin{equation}
\theta^* = \arg\min_{\theta} \mathcal{L}_{\text{total}}(\theta; x, t)
\end{equation}

Mariappan et al. \cite{mariappan2024learning} demonstrated that accurately predicting thermoacoustic instabilities depends heavily on optimizing the contributions of physics-informed loss components and fine-tuning hyperparameters.

\section{Current Challenges and Limitations}

Despite recent advances in the application of PINNs for combustion modeling, several critical challenges remain. These limitations hinder their full-scale deployment in practical combustion systems. Key issues include data quality, computational demands, generalization, multi-physics integration, and real-time applicability, as highlighted in recent studies \cite{song2024physics, sun2023physics, ji2021stiff, zhang2024crk, wu2023application, mariappan2024learning}.

\subsection{Data Quality and Availability}

Obtaining high-fidelity data for training PINNs is inherently difficult due to the extreme operating conditions in combustion systems. Experimental datasets often suffer from limited spatial and temporal resolution, introducing measurement uncertainty that propagates into PINNs predictions \cite{mariappan2024learning, song2024physics}. This uncertainty affects the data-driven component of the loss function:
\begin{equation}
\mathcal{L}_{\text{data}} = \frac{1}{N} \sum_{i=1}^N w_i (\hat{y}_i - y_i)^2
\end{equation}
and must be carefully managed to ensure predictive accuracy.

Additionally, as Wu et al. \cite{wu2023application} discussed, resolving key combustion features like flame fronts and scalar dissipation rates requires fine-grained data:
\begin{equation}
\Delta x_{\text{min}} \leq \min(l_K, l_F, l_{\chi})
\end{equation}
Inadequate resolution can impair the model's ability to capture important combustion dynamics.

\subsection{Generalization and Scalability}

While PINNs have demonstrated high accuracy within trained regimes, their generalization to unseen operating conditions remains a challenge. Predictive errors tend to increase when extrapolating beyond the training data distribution \cite{ji2021stiff, zhang2024crk}:
\begin{equation}
\varepsilon_{\text{gen}} = \|f_{\text{PINN}}(x_{\text{new}}) - f_{\text{true}}(x_{\text{new}})\|
\end{equation}

Scalability is another limitation when applying PINNs to large-scale combustion systems. The computational cost grows rapidly with the number of model parameters and the complexity of the physical phenomena being modeled \cite{sun2023physics}:
\begin{equation}
C_{\text{comp}} \propto \exp(\alpha N_{\text{params}})
\end{equation}

\subsection{Computational Demands and Training Complexity}

Training PINNs for stiff chemical kinetics and multi-scale combustion problems is computationally intensive. The total loss function typically incorporates multiple physical constraints:
\begin{equation}
\min_{\theta} \mathcal{L}_{\text{total}} = \mathcal{L}_{\text{data}} + \sum_{i=1}^M \lambda_i \mathcal{L}_{\text{physics},i}
\end{equation}

The training time scales with the number of epochs, data points, and enforced physics:
\begin{equation}
t_{\text{train}} \propto N_{\text{epochs}} \times N_{\text{samples}} \times N_{\text{constraints}}
\end{equation}

Additionally, PINNs often face gradient pathologies in stiff reaction systems, resulting in vanishing or exploding gradients \cite{nasiri2022reduced, ji2021stiff}:
\begin{equation}
\|\nabla_{\theta} \mathcal{L}\| \rightarrow 0 \quad \text{or} \quad \|\nabla_{\theta} \mathcal{L}\| \rightarrow \infty
\end{equation}

Reduced-order frameworks, such as Reduced-PINNs, have been proposed to alleviate these issues by transforming ODEs into integral formulations and reducing stiffness, as shown by Nasiri et al. \cite{nasiri2022reduced}.

\subsection{Integration with Experimental Systems}

For real-time applications, PINNs must deliver predictions on timescales much shorter than those of the physical process:
\begin{equation}
t_{\text{inference}} \ll t_{\text{process}}
\end{equation}

Mariappan et al. \cite{mariappan2024learning} emphasized the importance of integrating PINNs with high-frequency sensor data to enable adaptive control. Sensor sampling rates must comply with Nyquist criteria:
\begin{equation}
f_{\text{sample}} \geq 2f_{\text{max}}
\end{equation}

\subsection{Multi-Physics Coupling and Conflicting Constraints}

Complex combustion systems require simultaneous modeling of fluid dynamics, heat transfer, chemical kinetics, and sometimes acoustics. Coupling these multi-physics domains within a single PINNs framework introduces competing objectives in the loss terms \cite{song2024physics, wu2023application}:
\begin{equation}
\nabla_{\theta} \mathcal{L}_{\text{physics},i} \cdot \nabla_{\theta} \mathcal{L}_{\text{physics},j} < 0
\end{equation}

This conflict complicates optimization and can prevent convergence. The underlying PDEs often exhibit tight coupling, increasing numerical stiffness and model complexity:
\begin{equation}
\frac{\partial x_i}{\partial t} = f_i(x_1, \ldots, x_n, u) + g_i(x_1, \ldots, x_n)
\end{equation}

\subsection{Validation and Uncertainty Quantification}

Validating PINNs for industrial combustion applications is difficult due to limited high-fidelity experimental data. Quantifying uncertainties in predictions requires distinguishing among model, parameter, and data uncertainties \cite{mariappan2024learning, song2024physics}:
\begin{equation}
\sigma_{\text{pred}}^2 = \sigma_{\text{model}}^2 + \sigma_{\text{param}}^2 + \sigma_{\text{data}}^2
\end{equation}

Bayesian methods can provide posterior distributions of model parameters, offering a structured approach to uncertainty quantification:
\begin{equation}
P(y|x) = \int P(y|x, \theta) P(\theta|D) \, d\theta
\end{equation}

\subsection{Implementation and Practical Considerations}

Deploying PINNs in real-world combustion control systems requires seamless integration with existing architectures. The system response time must meet stringent requirements for dynamic control:
\begin{equation}
t_{\text{response}} = t_{\text{inference}} + t_{\text{processing}} + t_{\text{actuation}} < t_{\text{critical}}
\end{equation}

Additionally, model updates are necessary as new data becomes available, requiring continuous learning strategies:
\begin{equation}
\mathcal{M}_{\text{updated}} = f(\mathcal{M}_{\text{current}}, D_{\text{new}}, \theta_{\text{new}})
\end{equation}

Despite these challenges, PINNs continue to show promise. Ongoing research focuses on addressing computational inefficiencies, enhancing robustness, and improving generalization capabilities to extend the practical application of PINNs in combustion modeling \cite{ji2021stiff, song2024physics, wu2023application, mariappan2024learning}.

%%%%%%%%%%%%%%%%%%%%%%%%%%%%%%%%%%%%%%%%%%%%%
\section{Future Perspectives and Research Opportunities}

\subsection{Future Perspectives}

With ongoing advancements in computational power and algorithmic efficiency, PINNs are poised to play an increasingly important role in combustion modeling. Future research will focus on hybrid frameworks that integrate PINNs with conventional CFD solvers to address multi-physics and multi-scale phenomena more effectively. These hybrid models aim to combine the high-fidelity capabilities of CFD with the flexibility and data efficiency of PINNs.

One of the most promising areas is the prediction of complex turbulent flows in reacting systems. Accurately modeling turbulence-flame interactions remains a grand challenge in combustion science. PINNs offer a data-efficient approach to capture these nonlinear interactions, particularly in regimes where experimental data is sparse.

Another emerging direction involves the use of PINNs for chemical reaction mechanism reduction. By learning low-dimensional manifolds and identifying dominant species and reaction pathways, PINNs can facilitate the development of reduced-order models that maintain accuracy while enabling real-time simulations and control.

Further development will focus on incorporating uncertainty quantification (UQ) frameworks within PINNs to enhance robustness, especially under variable operating conditions and in the presence of noisy data. Applications in novel fuel systems such as hydrogen and ammonia combustion, as well as multi-fuel strategies, are expected to expand as PINNs frameworks evolve.

While traditional CFD will remain essential for validation and high-fidelity simulations, the intelligent integration of PINNs with CFD offers significant potential for improving the modeling, optimization, and control of clean combustion technologies \cite{song2024physics, wu2023application, mariappan2024learning}.

\subsection{Advanced Architecture Development}

Advancements in PINNs architecture are critical to improving their capacity to handle stiff chemical kinetics and multi-scale combustion processes. Zhang et al. \cite{zhang2024crk} introduced the CRK-PINNs framework, which enforces conservation laws to improve model generalization and eliminate non-physical solutions. Building on these concepts, future developments may incorporate adaptive neural network structures that dynamically allocate computational resources based on localized solution features.

Physics-guided attention mechanisms, as proposed by Song et al. \cite{song2024physics}, will likely play a role in focusing computational effort on critical regions such as flame fronts and reaction zones, enabling accurate capture of transient and localized events in combustion systems.

\subsection{Integration with Advanced AI Techniques}

The integration of PINNs with other AI methodologies, such as reinforcement learning (RL) and generative adversarial networks (GANs), is expected to accelerate progress in combustion modeling and control. Reinforcement learning, coupled with PINNs, can develop adaptive control strategies optimized for dynamic combustion environments \cite{mariappan2024learning}.

Meanwhile, physics-informed GANs can generate high-fidelity synthetic data to supplement limited experimental datasets, enhancing the training and performance of PINNs-based models \cite{song2024physics}.

\subsection{Real-time Monitoring and Control}

The deployment of PINNs in real-time monitoring and control systems will benefit from network compression techniques such as pruning and quantization \cite{mariappan2024learning}. These methods reduce computational demands, making real-time implementation feasible without sacrificing accuracy.

Edge computing architectures leveraging PINNs can facilitate distributed control strategies, where local controllers coordinate with centralized decision-making systems to maintain combustion stability and optimize performance.

\subsection{Multi-physics and Multi-scale Integration}

Future PINNs frameworks will advance toward unified architectures that incorporate tightly coupled physical processes, including turbulent flows, chemical reactions, heat transfer, and acoustic instabilities. This integration is necessary to capture the complex behavior of modern combustion systems accurately.

Additionally, multi-scale modeling approaches—incorporating asymptotic expansions or homogenization methods—will bridge micro- and macro-scale combustion phenomena, enabling comprehensive simulations of complex systems \cite{song2024physics}.

\subsection{Uncertainty Quantification and Robustness}

Quantifying uncertainties in PINNs-based models remains a key challenge. Future research will emphasize Bayesian PINNs and variational inference frameworks that propagate epistemic and aleatory uncertainties through complex combustion systems \cite{wu2023application}.

These approaches will provide probabilistic predictions, essential for risk assessment and robust control in practical combustion applications.

\subsection{Advanced Optimization Techniques}

Efficient training of PINNs, particularly for stiff reaction kinetics, requires novel optimization strategies. Future research will focus on physics-aware gradient conditioning and preconditioning techniques to enhance convergence and stability during training \cite{nasiri2022reduced, zhang2024crk}.

Multi-objective optimization frameworks balancing data fidelity, physical consistency, and model complexity will further improve PINNs' efficiency and generalization capabilities \cite{wu2023application}.

\subsection{Emerging Application Domains}

PINNs are expected to expand into emerging applications, such as hydrogen and ammonia-fueled combustion systems, which present additional challenges due to increased reaction stiffness and non-equilibrium transport phenomena \cite{song2024physics}. Addressing these challenges will require ongoing refinement of PINNs architectures and training methodologies.

\subsection{Integration with Quantum Computing}

The integration of PINNs with quantum computing represents a promising frontier for modeling complex combustion dynamics. Hybrid quantum-classical frameworks, such as Quantum PINNs (QPINNs), leverage quantum computational resources to enhance efficiency and scalability \cite{dehaghani2024hybrid, trahan2024quantum}.

Quantum-enhanced PINNs are designed to process high-dimensional, computationally intensive components—such as stiff ODE solvers—while classical networks manage other tasks. Figures~\ref{fig:qvinn_architecture} and~\ref{fig:cv_qvinn} illustrate QPINNs architectures applied to combustion modeling, demonstrating their potential to accelerate computations and enable real-time simulations.

\subsection{Environmental Impact Assessment}

Future PINNs applications will focus on environmental impact minimization, optimizing combustion processes to reduce emissions of CO$_2$, NO$_x$, and particulate matter. Life-cycle analysis frameworks integrated with PINNs will allow comprehensive assessments of sustainability and environmental performance throughout the combustion process lifecycle \cite{velmurugan2024enhancing, fedorov2023improving}.

\subsection{Cross-domain Applications}

Extending PINNs methodologies to other domains, such as energy systems, aerospace, and biomedical engineering, presents new opportunities for cross-disciplinary applications. Transfer learning and domain adaptation techniques will facilitate knowledge transfer between different physical systems, improving the generality and efficiency of PINNs-based models \cite{wang2025transfer, xu2023transfer}.

These cross-domain applications underscore the versatility of PINNs and their potential to advance scientific understanding and technological development in various fields.

%%%%%%%%%%%%%%%%%%%%%%%%%%%%%%%%%%%%%%%%%%
\section{Future Perspectives and Research Opportunities}

\subsection{Future Perspectives}

As computational power and algorithmic efficiency continue to advance, PINNs are poised to play an increasingly transformative role in combustion modeling. Future research is expected to focus on hybrid frameworks that integrate PINNs with conventional CFD solvers, enabling the efficient simulation of multi-physics and multi-scale combustion systems.

One key area of development is the prediction of complex turbulent flows in reactive systems, particularly the intricate interactions between turbulence and flame dynamics. Capturing these interactions remains a grand challenge in combustion modeling. PINNs offer the potential to improve prediction fidelity by leveraging limited datasets and Incorporating physics constraints in the learning process.

Another promising application of PINNs lies in chemical reaction mechanism reduction. By learning low-dimensional manifolds and identifying key species and reaction pathways, PINNs can facilitate the development of reduced-order chemical models. These models can significantly enhance computational efficiency, making real-time combustion simulations and control applications more feasible.

Additionally, integrating uncertainty quantification (UQ) within PINNs frameworks is a critical area of ongoing research. UQ can enhance the reliability and robustness of predictions, particularly under varying operating conditions and in the presence of noisy data. PINNs are also anticipated to expand into novel fuel systems, such as hydrogen and ammonia combustion, and multi-fuel strategies, which present unique modeling challenges due to their increased stiffness and non-equilibrium transport processes.

While traditional CFD methods remain indispensable for high-fidelity simulations and validation, the intelligent combination of PINNs and CFD promises significant advancements in the modeling, optimization, and control of clean combustion technologies \cite{song2024physics, wu2023application, mariappan2024learning}.

\subsection{Advanced Architecture Development}

Enhancing the architecture of PINNs is crucial for resolving stiff chemical kinetics and multi-scale combustion phenomena. Zhang et al. \cite{zhang2024crk} introduced the CRK-PINNs framework, which explicitly incorporates conservation laws, improving model generalization and addressing non-physical solutions. Building on these concepts, adaptive neural network architectures are being developed to dynamically adjust their structure based on localized solution gradients:
\begin{equation}
\mathcal{N}_{\text{adaptive}} = f(\nabla u, \text{ResNet}_{\theta}, \text{AttentionBlock}_{\phi}).
\end{equation}

Physics-guided attention mechanisms, as proposed by Song et al. \cite{song2024physics}, focus computational resources on regions of interest, such as flame fronts and reaction zones. This approach enhances the accuracy of localized and transient combustion event predictions:
\begin{equation}
\alpha(x,t) = \text{softmax}\left(\frac{Q_{\theta}K_{\phi}^T}{\sqrt{d_k}}\right)V_{\psi}.
\end{equation}

\subsection{Integration with Advanced AI Techniques}

Integrating PINNs with other artificial intelligence (AI) techniques offers substantial potential for advancing combustion modeling and control. Reinforcement learning (RL), when combined with PINNs, facilitates adaptive and optimal control strategies in dynamic combustion environments \cite{wu2023application, mariappan2024learning}:
\begin{equation}
\pi^*(s) = \arg\max_a \mathbb{E}[R_t \,|\, s_t = s,\, a_t = a,\, \mathcal{M}_{\text{PINN}}].
\end{equation}

Physics-informed Generative Adversarial Networks (GANs) can augment sparse datasets and improve PINN model training. By adding physics-consistent constraints to the GAN loss function, model generalization can be enhanced \cite{song2024physics}:
\begin{equation}
\min_G \max_D \, \mathbb{E}_{x \sim p_{\text{data}}}[\log D(x)] + \mathbb{E}_{z \sim p_z}[\log(1 - D(G(z)))] + \lambda \mathcal{L}_{\text{physics}}.
\end{equation}

\subsection{Real-Time Monitoring and Control}

Compressed PINNs models, utilizing techniques such as network pruning and quantization, are essential for real-time combustion monitoring and control. These techniques reduce computational demands while preserving predictive accuracy \cite{mariappan2024learning}:
\begin{equation}
\mathcal{M}_{\text{compressed}} = Q(\mathcal{M}_{\text{full}}, \epsilon_{\text{tol}}).
\end{equation}

Edge computing frameworks enable distributed and hierarchical control strategies, where localized PINN-based controllers coordinate with global optimization frameworks to maintain combustion system stability and efficiency:
\begin{equation}
u_i(t) = \mathcal{K}_i(\mathcal{M}_{\text{local}}, \mathcal{M}_{\text{global}}, x_i(t)).
\end{equation}

\subsection{Multi-Physics Integration}

Combustion systems exhibit tightly coupled multi-physics interactions involving turbulence, detailed chemistry, heat transfer, and acoustic instabilities. Future PINNs frameworks will integrate these interactions within unified architectures \cite{zhang2024crk, mariappan2024learning}:
\begin{equation}
\frac{\partial \mathbf{x}}{\partial t} = \mathcal{F}_{\text{PINN}}(\mathbf{x}, t, \{\mathcal{M}_i\}_{i=1}^N).
\end{equation}

Multi-scale modeling approaches will further bridge macro- and micro-scale combustion processes through embedded asymptotic expansions or homogenization techniques \cite{song2024physics}:
\begin{equation}
u(x,t) = u_{\text{macro}}(x,t) + \sum_{i=1}^N \epsilon_i u_{\text{micro},i}\left(\frac{x}{\epsilon_i}, \frac{t}{\epsilon_i^2}\right).
\end{equation}

\subsection{Uncertainty-Aware Frameworks}

Bayesian neural networks, integrated with physics-informed constraints, are being developed to quantify and propagate uncertainties in PINN-based combustion models \cite{mariappan2024learning, wu2023application}:
\begin{equation}
P(y|x,D) = \int P(y|x,\omega)P(\omega|D)d\omega.
\end{equation}

Physics-constrained variational inference (VI) methods offer robust uncertainty propagation in multi-physics PINNs, as shown in recent studies \cite{song2024physics}:
\begin{equation}
\mathcal{L}_{\text{VI}} = \text{KL}(q_{\phi}(\omega)||p(\omega)) - \mathbb{E}_{q_{\phi}(\omega)}[\log p(D|\omega)] + \lambda\mathcal{L}_{\text{physics}}.
\end{equation}

\subsection{Advanced Optimization Techniques}

Training PINNs for stiff reaction kinetics remains computationally demanding. Future research will explore physics-aware gradient conditioning and preconditioning techniques to improve convergence and stability \cite{nasiri2022reduced, zhang2024crk}:
\begin{equation}
\theta_{k+1} = \theta_k - \alpha_k\mathbf{H}_{\text{physics}}^{-1}\nabla_{\theta}\mathcal{L},
\end{equation}
where \( \mathbf{H}_{\text{physics}} \) is a physics-informed Hessian or preconditioner matrix.

Multi-objective optimization frameworks are also being developed to balance competing modeling objectives, including data fidelity, physics consistency, and model complexity \cite{wu2023application}:
\begin{equation}
\min_{\theta} \left\{ \mathcal{L}_{\text{data}}(\theta), \mathcal{L}_{\text{physics}}(\theta), \mathcal{L}_{\text{complexity}}(\theta) \right\}.
\end{equation}

\subsection{Emerging Application Domains}

PINNs are expanding into emerging application areas such as hydrogen combustion, ammonia-fueled systems, and multi-fuel combustion strategies \cite{song2024physics}. These systems introduce additional modeling complexities, including increased stiffness in reaction kinetics and non-equilibrium transport processes, requiring further advancements in PINNs frameworks.

\subsection{Integration with Quantum Computing}

The integration of PINNs with quantum computing represents an emerging frontier in combustion modeling. Hybrid quantum-classical frameworks leverage quantum computational dynamics to enhance the efficiency and capability of PINNs \cite{dehaghani2024hybrid, trahan2024quantum}.

\begin{figure}[htbp]
    \centering
    \includegraphics[width=0.8\textwidth]{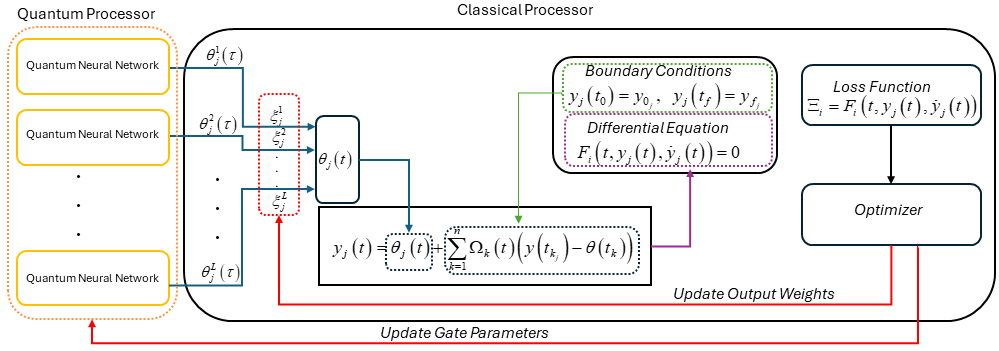}
    \caption{Schematic of the Quantum Physics-Informed Neural Network (QPINN) hybrid quantum-classical architecture. Quantum circuits process parameterized quantum states, with classical components handling data and optimization loops \cite{dehaghani2024hybrid}.}
    \label{fig:qvinn_architecture}
\end{figure}

One such architecture is the Quantum Physics-Informed Neural Network (QPINN), which combines quantum neural networks (QNNs) with classical components. In these systems, QNNs approximate solutions to partial differential equations (PDEs) relevant to combustion processes. Figure~\ref{fig:qvinn_architecture} presents a schematic of a hybrid QPINN architecture, where quantum circuits interface with classical optimization loops \cite{dehaghani2024hybrid}.

\begin{figure}[htbp]
    \centering
    \includegraphics[width=0.8\textwidth]{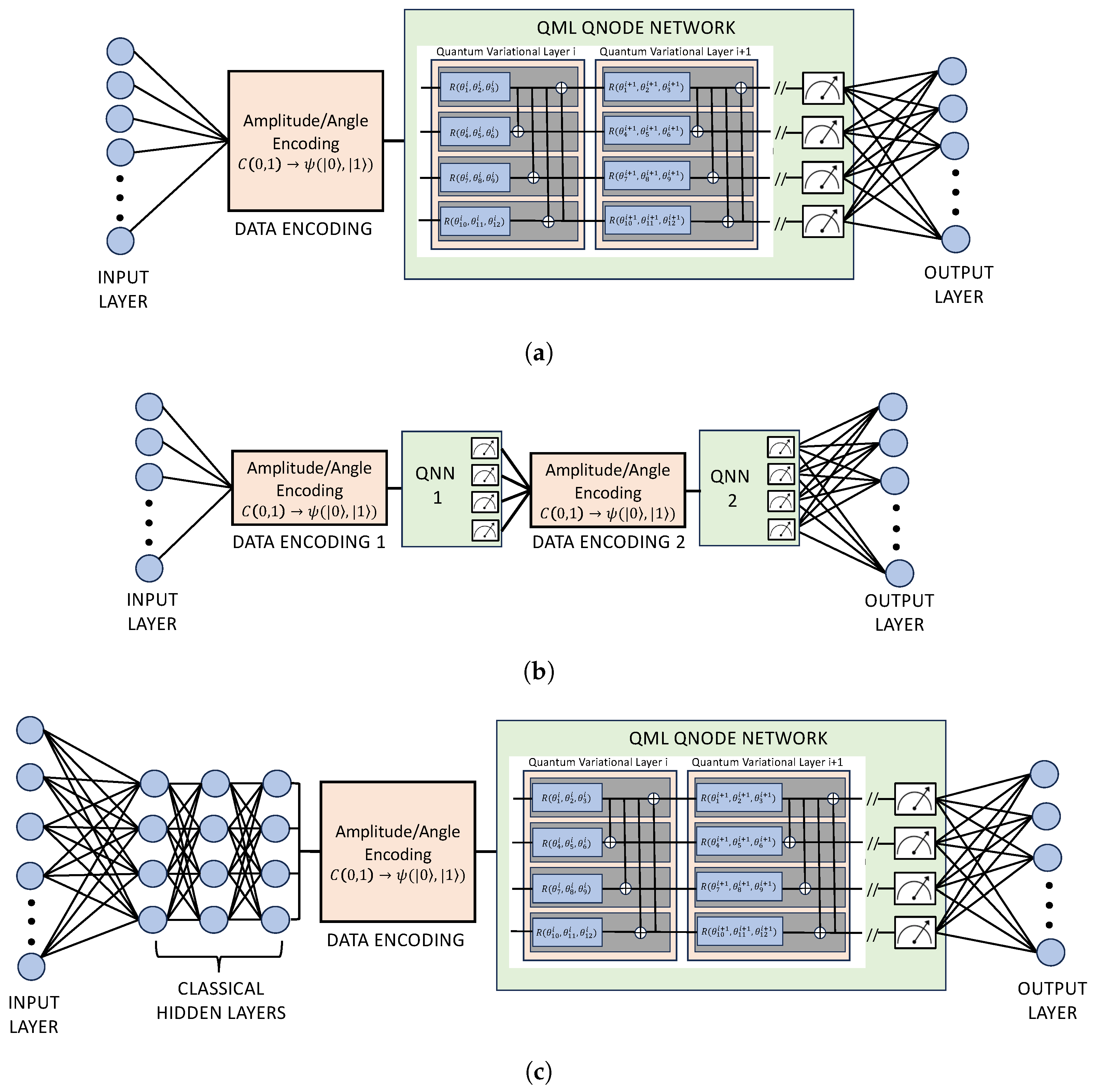}
    \caption{Continuous Variable (CV) QPINNs framework for solving PDEs using Gaussian and non-Gaussian quantum gates acting on qumodes. This approach demonstrates quantum circuit efficiency in approximating stiff ODEs and PDEs relevant to combustion modeling \cite{trahan2024quantum}.}
    \label{fig:cv_qvinn}
\end{figure}

Continuous Variable (CV) quantum computing frameworks further enhance this approach by employing Gaussian and non-Gaussian quantum gates to approximate solutions to stiff ODEs and PDEs, as shown in Figure~\ref{fig:cv_qvinn} \cite{trahan2024quantum}.

\subsection{Environmental Impact Assessment}

Future applications of PINNs will increasingly emphasize environmental impact optimization, aiming to minimize emissions such as CO$_2$ and NO$_x$, alongside operational costs:
\begin{equation}
    \min_{\theta} \{\text{CO}_2(u_{\theta}), \text{NO}_x(u_{\theta}), \text{Cost}(u_{\theta})\}.
\end{equation}

Integrating life-cycle analysis (LCA) enables comprehensive sustainability assessments by accounting for environmental impacts over the entire process duration \cite{velmurugan2024enhancing, fedorov2023improving}:
\begin{equation}
    \text{Impact}_{\text{total}} = \int_0^T \sum_{i=1}^N w_i \, \text{Impact}_i(u_{\theta}(t)) \, dt.
\end{equation}

\subsection{Cross-Domain Applications}

Extending PINNs methodologies to other domains can yield broader impacts. Transfer learning operators can adapt combustion models to different fields or operational conditions:
\begin{equation}
    \mathcal{T}: \mathcal{M}_{\text{source}} \rightarrow \mathcal{M}_{\text{target}}.
\end{equation}

Domain adaptation techniques can enhance model performance by minimizing discrepancies between different datasets \cite{wang2025transfer, xu2023transfer}:
\begin{equation}
    \mathcal{L}_{\text{transfer}} = \mathcal{L}_{\text{task}} + \lambda \mathcal{L}_{\text{domain}}.
\end{equation}

These emerging research directions underscore the transformative potential of PINNs in combustion science. By integrating advanced AI methods with physics-based principles, PINNs promise to enhance our understanding and control of complex combustion systems while promoting sustainable energy technologies.

\section{Conclusion}
PINNs have emerged as a transformative framework for overcoming the limitations of traditional combustion modeling. By integrating conservation laws, chemical kinetics, and turbulence models within neural network architectures, PINNs bridge the gap between data-driven machine learning and physics-based numerical methods. This integration enables the development of accurate, generalizable models, even in scenarios with sparse or incomplete data.

This review has presented a comprehensive overview of state-of-the-art PINNs applications in clean combustion, including flame dynamics, turbulent combustion, pollutant emissions prediction (NO\textsubscript{x}, soot, CO), and system optimization and control. Advances such as CRK-PINNs and DPINNs have shown significant progress in addressing stiff reaction kinetics and multi-scale challenges. Additionally, the incorporation of DeepONet architectures and hybrid quantum-classical PINNs paves the way for real-time control and monitoring. Integration with reinforcement learning and generative adversarial networks further expands the scope of adaptive control and data augmentation in practical combustion systems.

Despite their promise, PINNs face several challenges. High computational costs during training, particularly in multi-physics and high-dimensional applications, remain a barrier. Ensuring stability and convergence when modeling stiff reaction systems, along with the need for rigorous uncertainty quantification, are ongoing areas of research. The lack of standardized frameworks and domain-specific solutions also hinders broader adoption in industrial settings.

Looking forward, hybrid frameworks combining PINNs with traditional CFD methods are likely to offer the best balance of accuracy, interpretability, and efficiency. Advances in neural network architecture, multi-scale and multi-physics integration, uncertainty quantification, and physics-guided optimization will further enhance PINNs’ predictive capabilities. Expanding their application to novel fuels such as hydrogen and ammonia, and adopting transfer learning for cross-domain adaptability, will accelerate the development of clean, efficient, and low-emission combustion technologies.

\section{Declaration of generative AI and AI-assisted technologies in the writing process}
During the preparation of this work the authors used AI in order to improve the readability and language of the manuscript. After using this tool/service, the authors reviewed and edited the content as needed and take full responsibility for the content of the published article.
\bibliographystyle{unsrt}
\bibliography{references}

\end{document}